\documentclass[11pt,a4paper]{article} 
\pdfoutput=1
\usepackage{jheppub}
\usepackage{enumerate}
\usepackage{epsfig}
\usepackage{wasysym} 
\usepackage{mathrsfs} 
\usepackage{graphicx} 
\usepackage{amsfonts} 
\usepackage{amsbsy} 
\usepackage{amscd}
\usepackage{slashed} 
\usepackage{multirow} 
\usepackage{color}
\definecolor{myred}{rgb}{0.6,0,0} %usage:  {\textcolor{myred}{Hello World}}
\definecolor{myblue}{rgb}{0,0.2,0.4}
\definecolor{mygreen}{rgb}{0,0.9,0.1}
\definecolor{hc}{rgb}{.9,0.1,0.7}
\definecolor{hcout}{rgb}{.9,0.7,0.9}
\definecolor{Orange}{rgb}{1.,0.65,0.}
\usepackage[toc,page]{appendix}

\makeatletter

\newcommand{\fmslash}[2][0mu]{%
  \mathchoice
    {\fmsl@sh\displaystyle{#1}{#2}}%
    {\fmsl@sh\textstyle{#1}{#2}}%
    {\fmsl@sh\scriptstyle{#1}{#2}}%
    {\fmsl@sh\scriptscriptstyle{#1}{#2}}}
\newcommand{\fmsl@sh}[3]{%
  \m@th\ooalign{$\hfil#1\mkern#2/\hfil$\crcr$#1#3$}}
\makeatother

\newcommand{\lsim}{{\;\raise0.3ex\hbox{$<$\kern-0.75em\raise-1.1ex\hbox{$\sim$}}\;}}
\newcommand{\gsim}{{\;\raise0.3ex\hbox{$>$\kern-0.75em\raise-1.1ex\hbox{$\sim$}}\;}}

\newcommand{\met}{{\fmslash E_T}}

\usepackage{array}
\newcolumntype{C}[1]{>{\centering\arraybackslash$}p{#1}<{$}}

\usepackage{bm} % bold in math mode
\newcommand{\be}{\begin{equation}}
\newcommand{\ee}{\end{equation}}
\newcommand{\bes}{\begin{equation*}}
\newcommand{\ees}{\end{equation*}}
\newcommand{\bea}{\begin{eqnarray}}
\newcommand{\eea}{\end{eqnarray}}
\newcommand{\beas}{\begin{eqnarray*}}
\newcommand{\eeas}{\end{eqnarray*}}
\newcommand{\Tr}{\text{Tr}}

\title{Heavy stable charged tracks as signatures of non-thermal dark matter at the LHC : 
a study in some non-supersymmetric scenarios} 

\author[a]{Avirup Ghosh,}
\author[a]{Tanmoy Mondal,} 
\author[a]{Biswarup Mukhopadhyaya} 
  \affiliation[a]{Regional Centre for Accelerator-based Particle Physics,
Harish-Chandra Research Institute, HBNI,
Chhatnag Road, Jhunsi, Allahabad - 211 019, India} 
\emailAdd{avirupghosh@hri.res.in}
\emailAdd{tanmoymondal@hri.res.in}
\emailAdd{biswarup@hri.res.in}

\abstract{
We consider two theoretical scenarios, each including a $\mathbb{Z}_{2}$-odd sector
and leading to an elementary dark matter candidate. The first one is a variant
of the Type-III seesaw model where one lepton triplet is $\mathbb{Z}_{2}$-odd, together with
a heavy sterile neutrino. It leads to a fermionic dark matter, together with
the charged component of the triplet being a quasi-stable particle which decays
only via a higher-dimensional operator suppressed by a high scale. The second
model consists of an inert scalar doublet together with a $\mathbb{Z}_{2}$-odd right-handed
Majorana neutrino dark matter. A tiny Yukawa coupling delays the decay of the
charged component of the inert doublet into the dark matter candidate, making
the former long-lived on the se of collider detectors. The parameter space
of each model has been constrained by big-bang nucleosynthesis constraints,
and also by estimating the contribution to the relic density through freeze-out of
the long-lived charged particle as well the freeze-in production of the dark matter
candidate. We consider two kinds of signals at the Large Hadron Collider
for the first kind of models, namely {\it two charged tracks} and {\it single track +
$\met$}. For the second kind, the characteristic signals are {\it opposite} as
well as {\it same-sign charged track pairs}. We perform a detailed analysis using
event selection criteria consistent with the current experimental programmes.
It is found that the scenario with a lepton triplet can be probed upto
960(1190) GeV with an integrated luminosity of 300(3000) $fb^{-1}$, while the
corresponding numbers for the inert doublet scenario are 630(800) GeV. Furthermore,
the second kind of signal mentioned in each case allows us to differentiate 
different dark matter scenarios from each other.
}

\preprint{HRI-RECAPP-2017-007\\$\textrm{}$\hfill \today}

\begin{document}
\maketitle

\newpage

%------------------------------------------- INTRODUCTION -------------------------------------------------
\section{Introduction} \label{sec:intro}
The presence of dark matter (DM), comprising about 23.8\% of the
energy density of the universe, is practically undeniable today\cite{Ade:2015xua,Kolb:1990vq}.
It is also largely felt, due to observations such as the bending
of light around the tail of bullet clusters, that at least a
substantial fraction of DM consists of some yet unknown elementary
particle(s) possessing neither electric charge nor color. The
stability of any such massive particle needs to be further justified
in a theoretical framework which needs to go beyond the standard model
(SM). A large number of scenarios have thus been proposed and
explored. Supersymmetric (SUSY) theories are rather strong contenders
in this context; with baryon-and lepton-number conserved (at least to
odd units), SUSY automatically offers a $\mathbb{Z}_2$ symmetry that makes the
lightest SUSY particle stable and a suitable DM candidate. However,
the presence of SUSY around the TeV-scale also implies the existence
of an additional colored sector consisting of squarks and gluinos,
amenable to production at a machine such as the Large Hadron Collider
(LHC). Non-observation of the resulting signals so far, while not
necessarily writing off SUSY as a possibility, keeps interests alive
in many other models where the SM is extended in the electroweak
sector alone.

Other than predicting a non-negligible part of the observed relic
density, the terrestrial observation of a DM candidate particle can
come through (a) recoil events in direct search experiments
set up world-wide and (b) collider events with large missing
energy/momentum. Both of these, however, are contingent upon the DM
candidate having a minimum interaction strength with SM
particles. Depending on the theoretical scenario, one may, for
example, envision situations where a DM candidate yielding the right
relic density and giving rise to missing transverse energy (MET)
signals at the LHC may yield far too small a recoil rate for detection
at direct search experiments. An example of this is a keV-scale
gravitino which passes off as `warm dark mater' candidate\cite{Covi:2007xj,Covi:2009bk,Co:2016fln,
Ibe:2010ym,Shoemaker:2009kg,Viel:2005qj,Baltz:2001rq,Kawasaki:1996wc,Fukai:1985az}.

Somewhat more remarkable are situations where the DM candidate is far
too feebly interacting for all heavier particles to decay into it with
noticeable rates within the periphery of collider detectors. One does
not have any events with MET in such situations. On the other hand,
the next heavier particle which has the DM candidate in the final
state as the only channel of its decay becomes stable, or at any rate,
long-lived on the scale of collider detectors\footnote{This is also possible 
if the dark matter candidate is closely degenerate with a charged particle in 
the `dark sector', as discussed, for example, in~\cite{Khoze:2017ixx}.}.
And the characteristic signal of such a scenario turns out to be highly 
ionizing charged tracks bearing the footprints of heavy particles, which can 
be noticed in both the tracking chambers and the muon detector.

The most obvious example of this is a SUSY scenario with 
right-chiral neutrino superfields added to the spectrum of the minimal
SUSY standard model (MSSM), with, say, just Dirac masses for neutrinos. 
In such a case, one of the right-chiral sneutrinos which have no
interaction excepting extremely small Yukawa couplings ($\simeq
10^{-13}$) becomes a strong contender for the role of the DM candidate
\cite{Gupta:2007ui,Biswas:2009rba,Biswas:2009zp,Heisig:2011dr,Heisig:2012zq,
Heisig:2013rya,Heisig:2015yla,Banerjee:2016uyt}.
This is especially true if the scalar mass parameters have a
common origin at a high scale, as the right-sneutrino masses evolve
negligibly while all other scalars are boosted through gauge
interactions as one comes down to the TeV-scale.  This results in such
sneutrino states being the lightest SUSY particle (LSP) over a
substantial region of the parameter space. It is also often likely in
such cases for the lighter stau $\tilde{\tau}_1$ to be the
next-to-lightest SUSY particle (NLSP), being lighter than the lightest
neutralino ($\chi^0_1$).  All SUSY cascades, and also direct
Drell-Yan production, at the LHC should then lead to the production of
stau-pairs whose decays into the (right) sneutrino LSP is an
excruciatingly slow process, practically never seen within the detector.
Thus one has novel SUSY (and dark matter) signals consisting in {\it
not MET but stable charged tracks} in the collider detectors.

There can of course be other kinds of signals for such scenarios. 
First, one can have a long-lived coloured particle instead, which 
hadronises within the detector. The signal, or at any rate, the probability
of such stable charged tracks, then depends on hither to unknown 
fragmentation functions. In such cases, one may witness disappearing 
tracks, displaced vertices or even no visible tracks at all. While 
acknowledging such possibilities, we devote the present paper to the 
discussion of scenarios with charged tracks that are stable on the scale
of the detector.

A DM candidate of this nature is of course unable to thermalise with
the cosmic soup, and thus its contribution to the relic density is not
obtainable by solving the Boltzmann equation as in the case of thermal
DM particles.  On the other hand, if one neglects non-thermal
production and assumes that they are produced in the universe only via
the decay of the NLSP (which freezes out before decaying), then the
latter's contribution to the relic density can be scaled appropriately
to obtain at least an approximate upper limit on the mass of the
non-thermal DM particle \cite{Feng:2003xh,Feng:2004we}. Moreover, it is 
desirable for the NLSP to
have a lifetime not exceeding about 100 seconds, if the observed
abundance of light elements has to be commensurate with Big-Bang
Nucleosynthesis (BBN) in standard cosmology
\cite{Jittoh:2011ni,Kusakabe:2010cb,Kusakabe:2009jt,Holtmann:1998gd}.
Sufficiently large
regions consistent with these as well as collider and low-energy
constraints have been found allowed both in the (MSSM +
right-neutrinos) scenario and its counterpart in constrained MSSM
(CMSSM) based on minimal supergravity (mSUGRA)\cite{Banerjee:2016uyt}.

The collider signals of such stable charged tracks are rather
conspicuous in general. However, one needs to differentiate them from
muons which, too, leave their mark in both the tracker and the muon
chamber. Detailed theoretical studies on the merits of various event
selection criteria (which also need to address cosmic ray backgrounds)
have taken place, side by side with various cuts actually used by the
experimental collaborations. In general, it is found that the stable
charged particles of the aforementioned kind carry much higher $p_T$
than muons, a feature that is potentially an excellent discriminator.
However, when the mass of the stable particle is on the higher side,
(about 500 GeV or higher), it is more efficient from the standpoint of
statistical significance to use the velocity $\beta$ as measured
from the time delay between the inner tracker and the muon
chamber. Additional criteria such as the rate of energy loss of the
charged object can, expectedly, buttress the selection criteria.

These discussions generally fit in rather appropriately into a SUSY
scenario where one not only has a stable R-odd DM candidate but also
some additional R-odd charged particle like a stau just above it in
mass. However, given the fact that we are yet to see any signature of
the strongly interacting superparticles at the LHC, it is desirable to
explore theoretical possibilities where the DM candidate arises via
augmentation of just the electroweak sector, but is again very feeble
in its interactions with other particles due to some characteristic
feature of the model. Two such models are discussed in this paper,
where stable charged tracks may occur at the LHC through the
production of some particle that decay into the DM candidate, but only
outside the detector. We have a spin-1/2 DM, produced upon the decay
of a charged fermion, in one of these illustrative cases. In the
other, the $\mathbb{Z}_2$-odd sector consists of an inert scalar doublet in
addition to a heavy right-handed Majorana neutrino dark matter. We
show in the next few  Sections how one expects signals of both these
scenarios in the form of heavy charged tracks. In addition, the
special characteristics of the individual models are reflected in some
additional observations. These are, for example, the number of single
charged track events vs that of a pair of charged tracks, or same-sign
vs opposite-sign charge tracks. We emphasize that such observations
enable one to find out the actual nature of the new physics scenario
by analyzing the stable charged track signals.

Since we illustrate our point with two disparate scenarios, a little
extra care needs to be taken in deriving the constraints obtained from
the frozen-out quasi-stable (charged) particle density scaled appropriately.
If the decay width of such particles is such that a non-negligible DM density
is created even before the freeze-out of the former, then this latter, too,
contribute to the relic. In this work we have included the contribution of these
`frozen-in' DM particles, over and above that from the quasi-stable particles
which freeze out. This inclusion was of little consequences in \cite{Banerjee:2016uyt}
where the freeze-in effect is found to be rather small in view of the very
small coupling strengths of NLSP to the LSP.

It is important to identify regions in the parameter space
of each relevant model, where  signals of the above kinds can
be observed.  Keeping this in mind, we obtain
the regions where the lifetime of the quasi-stable charged particle,
while being less than 100~sec, ensure decays  outside the detector,
and is consistent with relic density bounds following the 
constraints stated above. This is in essence the space spanned by 
the mass difference between the quasi-stable particle and 
the DM candidate and the coupling pertinent to the decay of the former
 \footnote{There can, in principle, 
also be regions where the next-to-lightest (charged) particle decays
within the detector, thus leading to signals with disappearing
tracks. Such signals are not considered in this study and will be considered
subsequently.}.

Organization of the paper goes as follows: Section~\ref{sec:model} contains 
a brief description of the models and also various constraints leading to the 
feebly interacting DM candidates. Strategies for LHC-based analyses, including
those directed at minimising backgrounds, are incorporated in Section~\ref{sec:analysis}.
Section~\ref{sec:result} contains our numerical results and an account of the
discovery potential for such scenario. We summarize 
and conclude in  Section~\ref{sec:Summary and conclusions}. 

%***********************************************************************
%                   Models under consideration
%***********************************************************************
\section{Models and constraints}\label{sec:model}
 In this  Section we outline two(non-supersymmetric) new
 physics scenarios. A quasi-stable charged particle is envisioned 
 in each of them, which decays very slowly into DM particle. We also
 mention the constraints to which each model is subjected.
\subsection{Type III seesaw with sterile neutrino}
We consider, in addition to the SM particles, three fermionic SU(2)
triplets $\Sigma_{jR}$ of zero hypercharge, each composed of 
three right-handed Weyl Spinors of zero U(1)
hypercharge. Each $\Sigma_{jR}$ has the components 
($\Sigma^1_{jR}, \Sigma^2_{jR}, \Sigma^3_{jR}$) .   
Out of them one can construct the charged and neutral
triplets $(\Sigma_{jR}^+,\Sigma_{jR}^0,\Sigma_{jR}^-)$ where (j = 1-3),
represented by the 2$\times$2 matrix
\be\label{eq:sigma-bi-doublet}
\Sigma_{jR} = 
 \begin{bmatrix}
 \Sigma_{jR}^0/\sqrt{2} & \Sigma_{jR}^+ \\
 \Sigma_{jR}^- & -\Sigma_{jR}^0/\sqrt{2}
 \end{bmatrix},
\ee
where the fields  $(\Sigma_{jR}^+,\Sigma_{jR}^0,\Sigma_{jR}^-)$ have 
been defined as
\begin{center}
$\Sigma_{jR}^+ = \frac{1}{\sqrt{2}}(\Sigma^1_{jR} - i\Sigma^2_{jR})$,
$\Sigma_{jR}^- = \frac{1}{\sqrt{2}}(\Sigma^1_{jR} + i\Sigma^2_{jR})$,
$\Sigma_{jR}^0 = \Sigma^3_{jR}$,
\end{center}

 In addition, we consider a $\mathbb{Z}_2$ symmetry to ensure the
 stability of DM, under which the SM fields as well as two of the
 fermionic triplets are even. These fields are free to mix amongst
 themselves. Thus one generates two tree-level neutrino masses through
 the Type III seesaw mechanism and hence explains the observed
 mass-squared differences as suggested by neutrino oscillation
 experiments. On the other hand,the remaining triplet does not
 contribute to neutrino mass generation, because it is {\it odd under
 imposed $\mathbb{Z}_2$ symmetry}. The neutral component of the
 $\mathbb{Z}_2$-odd triplet mixes with a $\mathbb{Z}_2$-odd singlet
 sterile neutrino $\nu_{sR}$ (another right-handed Weyl fermion) to
 produce a dark matter candidate. If $\nu_{sR}$ be light enough compared
 to $\Sigma_{3R}$ and its mixing with $\Sigma_{3R}^0$ be small
 enough, the $\nu_{sR}$-dominated mass eigenstate can be a viable
 SuperWIMP Dark Matter candidate.

 Thus, over and above the SM part, the Lagrangian contains
 the following renormalizable terms \cite{Abada:2007ux,Abada:2008ea,Biggio:2011ja} 
 (written in terms of Weyl spinors):
\begin{eqnarray}
  \label{eq:lag_ren}
 \mathcal{L} &=& \Tr\left[\bar{\Sigma}_{jR}i
   \slashed{D}\Sigma_{jR}\right]-\frac{1}{2}\Tr\left[\bar{\Sigma}_{jR}M_{\Sigma}\Sigma_{jR}^{c}+h.c\right]-
 \left(\sqrt{2} \bar{L}_{Lj}Y_{\Sigma}\Sigma_{\alpha\!R}\tilde{\Phi} +
 h.c\right) \nonumber \\
 && + \frac{i}{2} \bar{\nu}_{sR}\slashed{\partial}\nu_{sR} -
 \frac{1}{2}\left(\bar{\nu}_{sR}M_{\nu_{s}}\nu_{sR}^{c} + h.c\right),
\end{eqnarray}
where $L_{L} \equiv (\nu_{L},l_{L})^{T}$, $\Phi \equiv (\phi^+ ,
(v+H+i\phi_{0})/\sqrt{2})^{T}$, $\tilde{\Phi}=
i\tau_{2}\Phi$, $\Sigma_{jR}^{c}= (\Sigma^{c})_{jL}=C\bar{\Sigma}_{jR}^{T}$ and summation over j and
$\alpha$ are implied. One has j = 1, 2, 3 and 
$\alpha$ = 1, 2, denoting generation
indices for the SM and triplet fermions, respectively,
involved in interactions with the Higgs doublet. It should be noted that in
eqn.(\ref{eq:lag_ren}), the Yukawa coupling terms for $\mathbb{Z}_2$-odd triplet
$\Sigma_{3R}$ as well as  sterile neutrino $\nu_{sR}$ is prohibited due to the
$\mathbb{Z}_2$ symmetry. As the hypercharges of $\Sigma_{3R}$ and $\nu_{sR}$
are both zero and in addition $T_3$ = 0 for $\Sigma_{3R}$, they have no Z-interaction,
thus evading direct search constraints on a DM candidate potentially emerging out of them.

 The smallness of $\nu_{sR} - \Sigma_{3R}^0$ mixing can be justified using
dimension-five interaction terms. One may assume that such
terms are artifacts of some new physics at a higher scale $\Lambda$, encapsulated in
the effective Lagrangian\cite{Chaudhuri:2015pna}
\begin{equation}
 \mathcal{L}_5 = \left(\frac{\alpha_{\Sigma\nu_{s}}}{\Lambda} \Phi^{\dagger}\bar{\Sigma}_{3R}\Phi\nu_{sR}^{c} +
 \frac{\alpha_{\Sigma\nu_{s}}}{\Lambda} \Phi^{\dagger}\bar{\Sigma}_{3R}^{c}\Phi\nu_{sR} +\frac{\alpha_{\nu_s}}
 {\Lambda}\Phi^{\dagger}\Phi\bar{\nu}_{sR}\nu_{sR}^{c} +\frac{\alpha_{\Sigma}}{\Lambda}\Phi^{\dagger}\bar{\Sigma}_{3R}\Sigma_{3R}^{c}\Phi\right)+ h.c.,
\end{equation}

though the various Wilson coefficients ($\alpha_{\Sigma\nu_{s}},
\alpha_{\Sigma},\alpha_{\nu_{s}}$) shown above are formally mentioned
in  the discussion that follows they have been all set to unity in our 
numerical calculation, keeping $\Lambda$ as the single parameter 
characterising all dimension-5 terms. This simplification does not affect
our results qualitatively.

The fields in the triplet-singlet sector in the four-component notation 
include the charged Dirac fermions
\begin{center}
$ \eta^{-}_j = \Sigma_{jR}^- + \Sigma_{jR}^{+c} $ ,$\;$  
$ \eta^{+}_j = \Sigma_{jR}^{-c} + \Sigma_{jR}^{+} $,
\end{center}
which have a definite mass ($M_{\Sigma}-\frac{\alpha_{\Sigma}v^{2}}{2\Lambda}$) for $j$ = 3. 
One also has in this sector  the Majorana Fermions
\begin{center}
$ \eta^{0}_j = \Sigma_{jR}^0 + \Sigma_{jR}^{0c} $ ,$\;$
$  N^{0} = \nu_{sR}^0 + \nu_{sR}^{0c} $.
\end{center}
The triplets of the first two families (corresponding to j = 1,2)
can of course mix with the SM leptons once electroweak 
symmetry is broken.

In terms of the Dirac and Majorana fermions, eqn.(\ref{eq:lag_ren})
can be rewritten (in terms of the individual components of SU(2)
doublets and triplets) as
\begin{eqnarray}\label{eq:lag_ren_extended}
\mathcal{L} &=& \bar{\eta_j}i\slashed{\partial}\eta_j + \frac{1}{2}\bar{\eta_j}^{0}i\slashed{\partial}\eta^{0}_j
- \bar{\eta_j}M_{\Sigma}\eta_j - \frac{1}{2}\bar{\eta}^{0}_jM_{\Sigma}\eta^{0}_j + g(\bar{\eta}^{0}_jW^{+}_{\mu}\gamma^{\mu}\eta_j + h.c)
- g\bar{\eta_j}W^3_{\mu}\gamma^{\mu}\eta_j \nonumber\\
&&  - [\Phi_0 \bar{\eta}^{0}_{\alpha}Y_{\Sigma}\nu_{Lj} + \sqrt{2} \Phi_0 \bar{\eta}_{\alpha}Y_{\Sigma}l_{Lj} +
\phi^+ \bar{\eta}^0_{\alpha}Y_{\Sigma}l_{Lj} - \sqrt{2} \phi^+ \bar{\nu}_{Lj}^c Y_{\Sigma}\eta_\alpha + h.c.] \nonumber \\
&& + \frac{i}{2}\bar{N}^0i\slashed{\partial}N^{0} - \frac{1}{2}\bar{N}^0 M_{\nu_s} N^0,
\end{eqnarray}
while the dimension-5 terms are,
\begin{eqnarray}
\mathcal{L}_5 &=& \frac{\alpha_{\Sigma\nu_{s}}}{\Lambda}(\frac{1}{\sqrt{2}} \phi^- \phi^+ \bar{\eta}^0_{3} N^0 + 
\phi^- \Phi_0 \bar{\eta}_{3} N^0 + \phi^+ \Phi_0^* \bar{N}^0 \eta_{3} + \frac{1}{\sqrt{2}}\Phi_0^* \Phi_0 \bar{N}^0 \eta^0_3) + h.c. \nonumber \\
&& (\phi^+ \phi^- + \Phi_0 \Phi_0^*)[\frac{\alpha_{\nu_s}}{\Lambda} \bar{N}^0 N^0 + \frac{\alpha_{\Sigma}}{\Lambda}
(\frac{1}{2} \bar{\eta}^0_3 \eta^0_3 + \bar{\eta}_3 \eta_3)],
\end{eqnarray}
where $\Phi_0 = (v+H+i\phi_{0})/\sqrt{2}$, the neutral component of the SM scalar doublet.

The $ N^{0}-\eta^{0}_3$ mass matrix is,
\begin{center}
$\begin{bmatrix}
 M_{\nu_s}- \frac{\alpha_{\nu_s}v^2}{\Lambda}  & \frac{\alpha_{\Sigma\nu_{s}}v^2}{\sqrt{2}\Lambda}  \\
 \frac{\alpha_{\Sigma\nu_{s}}v^2}{\sqrt{2}\Lambda} & 
 M_{\Sigma}-\frac{\alpha_{\Sigma}v^2}{2\Lambda}
 \end{bmatrix}$,
\end{center} 
  which, when diagonalized, yields the following  mass eigenstates,
\begin{equation}
\chi = \cos\beta\; N^{0}  - \sin\beta\; \eta^{0}_3 ,
\end{equation}
\begin{equation}
\psi = \sin\beta\; N^{0} + \cos\beta\; \eta^{0}_3 ,
\end{equation}
 where $\chi$ is the lighter state with mass,
 \begin{eqnarray}
  M_{\chi}  &=&\frac{1}{2}((M_{\nu_s}- \alpha_{\nu_s}v^2/\Lambda +  M_{\Sigma}-\alpha_{\Sigma}v^2/2\Lambda )^2 \nonumber \\
  && -\sqrt{(M_{\Sigma}-\alpha_{\Sigma}v^2/2\Lambda - M_{\nu_s} + \alpha_{\nu_s}v^2/\Lambda)^2 + 4 (\alpha_{\Sigma\nu_{s}}v^2/\Lambda)^2/2}), 
 \end{eqnarray}
 and $\beta$ is the mixing angle given by
 \begin{equation}
   \label{eq:mixing_beta}
  \tan2\beta =\dfrac{(\alpha_{\Sigma\nu_{s}}v^2)/\sqrt{2}\Lambda}{(M_{\Sigma}-\alpha_{\Sigma}v^2/2\Lambda - M_{\nu_s} + \alpha_{\nu_s}v^2/\Lambda )}.
 \end{equation}

 If we consider the new physics scale $\Lambda$ to be high enough, being on the 
 order of $10^{14}$ GeV or above, the dimension-5
 couplings become very small and hence $\chi$ interacts very weakly with the
 rest of the particles in the spectrum. One can safely assume that
 $\chi$ has never been in thermal equilibrium with the thermal soup
 during the evolution of the universe and hence is a viable candidate for 
 {\it SuperWIMP} (non-thermal) dark matter.  In such a scenario, $\chi$ may be 
 produced from the decay of next-to-lightest odd particle(s) (NLOP) 
 viz., $\eta^+_3$, $\eta^-_3$ and $\psi$. The discussion that follows
depends on the NLOP $\eta^{\pm}_3$ being effectively degenerate with the 
state $\psi$, something that is responsible for its quasi-stable character.
This may in principle be threatened by electromagnetic radiative 
corrections raising the $\eta^{\pm}_3$ mass \cite{Cirelli:2005uq}. 
Such a prospect can be alleviated by allowing the possibility of further mixing between $\eta^{\pm}_3$
  and some additional $\mathbb{Z}_2$-odd fermion(s) as outlined in appendix~\ref{app:charged-mixing}.
% {\bf This 
% may be alleviated by (a) allowing further mixing between $\eta^{\pm}_3$
% and some additional $\mathbb{Z}_2$-odd fermion(s), or (b) by postulating a 
% gradation of new physics scales, whereby a higher-dimension operator
% (here the one proportional to $\alpha_{\Sigma}$)with some suppression 
% scale may offset the electromagnetic correction. In the rest of the present 
% study, we do not go into the details of such modeling and treat the 
% near-degeneracy of $\eta^{\pm}_3$ and $\psi$ phenomenologically.}

 Initially the NLOPs were in thermal 
 equilibrium and eventually have frozen out at some point. The density of $\chi$ 
 therefore rises via the freeze-in process when the NLOP were still in thermal 
 equilibrium, and also through the density of latter dwindling via decay into
 $\chi$ after freeze-out. Thus, in order to compute the relic 
 density, we have to estimate the net abundance of $\chi$ generated both before and after the 
 freeze-out of the NLOP. When the NLOP were in thermal equilibrium, the \textit{freeze-in 
   yield} $Y_{\chi}$ of the DM is calculated using~\cite{Hall:2009bx},
 \begin{equation}\label{eq:freeze-in-yield}
 Y_{\chi} = \frac{45}{(4\pi^{4}) 1.66}\frac{g_{\Sigma}\; M_{Pl}\;\Gamma}{M_{\Sigma}^{2}~ h_{eff}\sqrt{g_{eff}}}\int^{x_f}_{x=0}K_1(x)x^3 dx,
 \end{equation} 
 where $g_{\Sigma}$ is the number of degrees of freedom of the NLOP, $M_{Pl}$ is the Planck 
 mass and $K_n(x)$ is the $n$th order modified Bessel function of second kind, 
 $x = \frac{M_{\Sigma}}{T}$. The $g_{eff}$ and $h_{eff}$ are given
 by the expressions for energy density, $\rho = g_{eff}(T)\frac{\pi^2}{30} T^4$ and 
 entropy density $ s = h_{eff}(T)\frac{2 \pi^2}{45} T^3$ respectively. 
 
 The yield of the DM  after freeze-out of NLOP is calculated by solving the coupled 
 Boltzmann equations~\cite{Gondolo:1990dk,Biswas:2016bfo},
 \begin{eqnarray}\label{eq:nlop-decay-yield}
 \frac{dY_{NLOP}}{dx} &=& -\sqrt{\frac{\pi}{45G}}\frac{g_{*}^{1/2}M_{\Sigma}}{x^2}\langle \sigma v_{Mol} \rangle(Y^{2}_{NLOP}-Y^{eq 2}_{NLOP})
                       - \sqrt{\frac{45}{\pi^{3}G}} \frac{x}{2 \sqrt{g_{eff}} M_{\Sigma}^{2}} \langle\Gamma\rangle Y_{NLOP},\nonumber\\
% \end{eqnarray} 
%  and
%  \begin{equation} 
 \frac{dY_{\chi}}{dx} &=& \sqrt{\frac{45}{\pi^{3}G}} \frac{x}{2 \sqrt{g_{eff}} M_{\Sigma}^{2}} \langle\,\Gamma\,\rangle Y_{NLOP} ,
%  \end{equation}
 \end{eqnarray} 
 where $Y_{NLOP}$ and $Y_{\chi}$ are the yield of NLOP and DM respectively.  
 The parameter $g_{*}$ is defined as the effective degrees of freedom of all the relativistic
 species still in thermal equilibrium when the NLOP freezes out.
  $\langle\,\Gamma\,\rangle$ is the thermally averaged decay 
 width of NLOP into DM and $G$ is the gravitational constant. The yield is related to 
 relic density via the relation,~\cite{Gondolo:1997km} 
 \begin{equation}
\Omega_{\chi}h^2 \simeq 2.8\times10^8\times\left(\frac{M_{\nu_s}}{GeV}\right)Y_{\chi}(x\to\infty),
 \end{equation}

 \begin{figure}[t]
 \begin{center}
 \includegraphics[width=7.5cm]{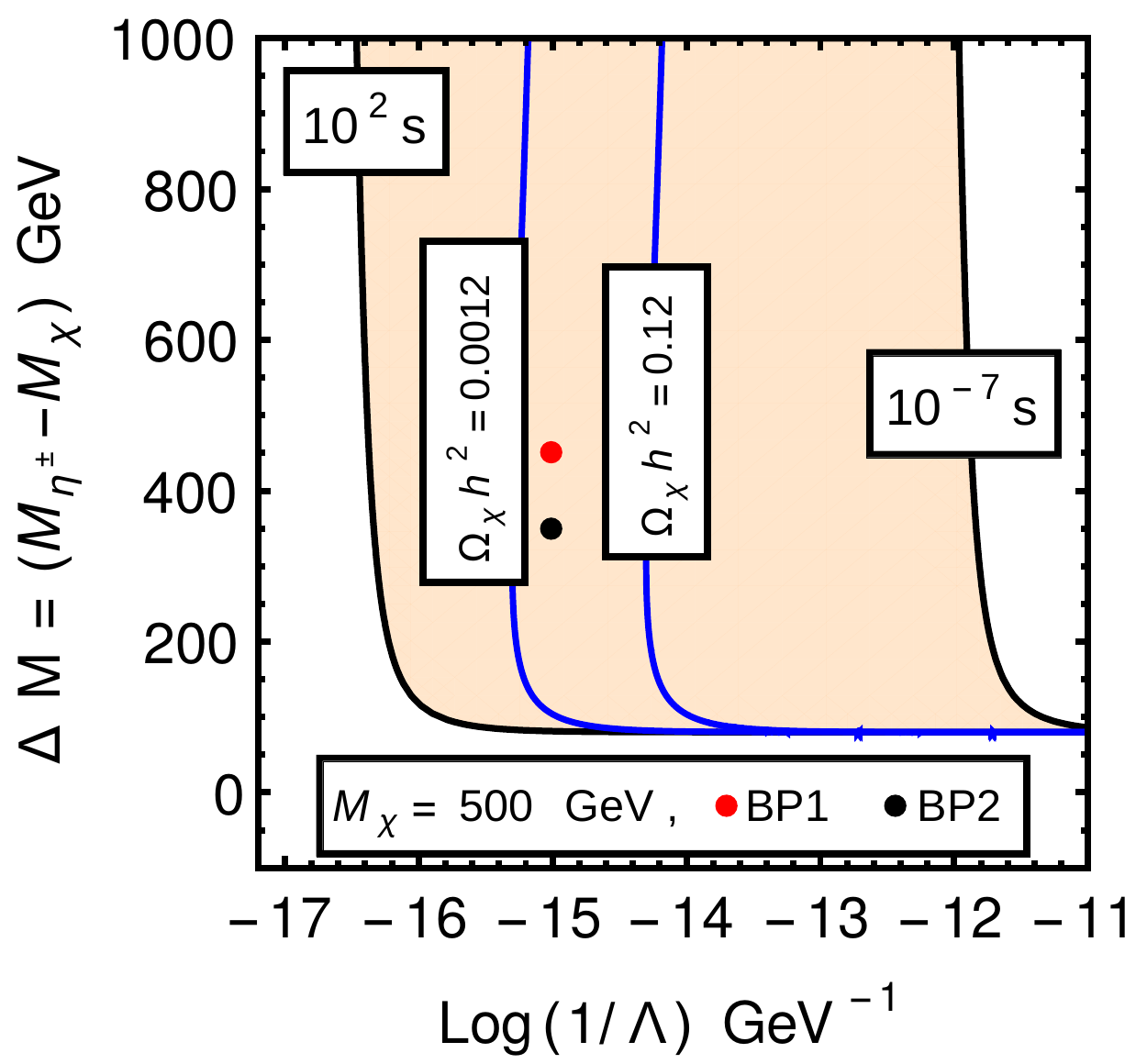}
 \includegraphics[width=7.5cm]{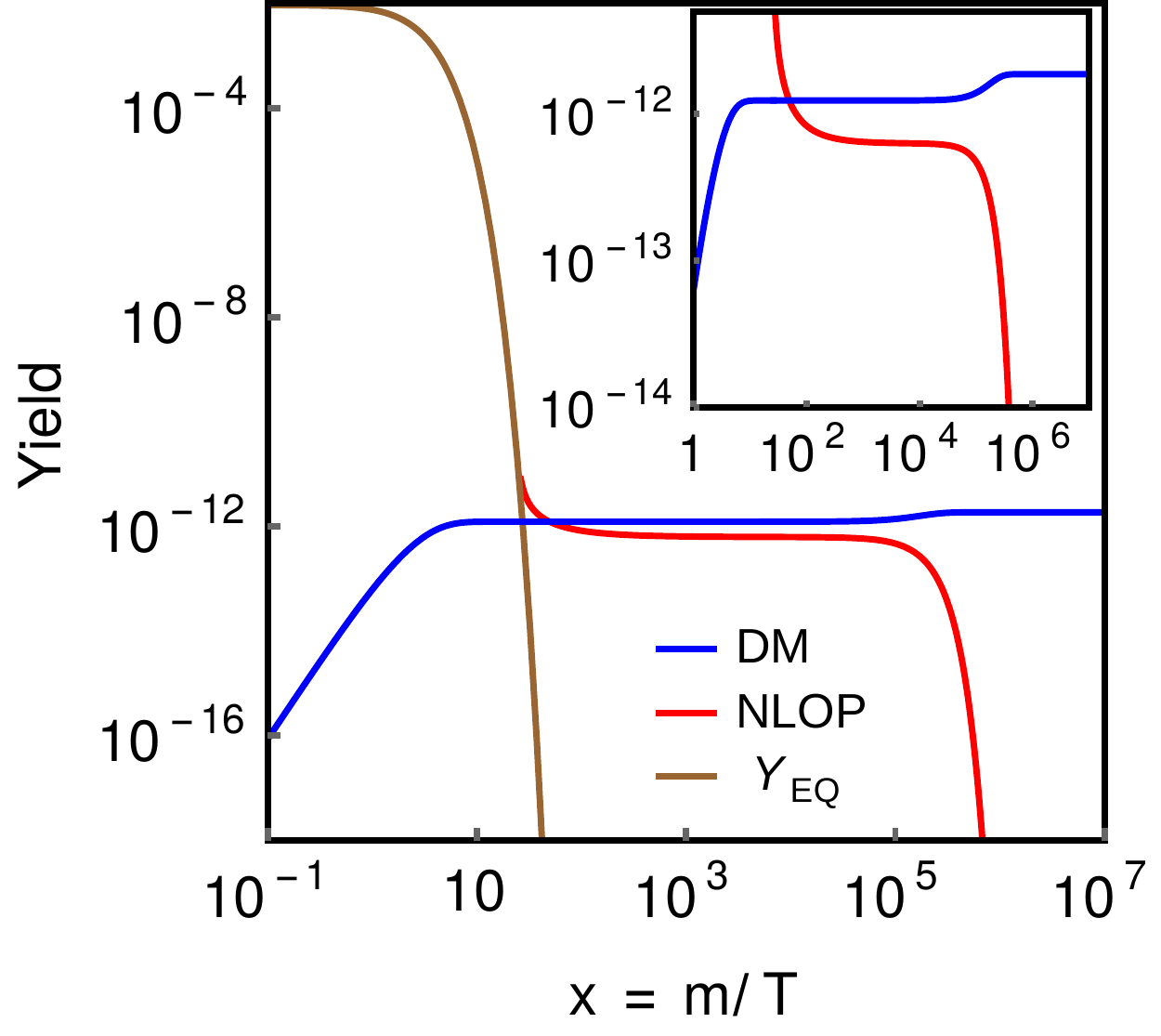}
 \caption{Contours of life time of HSCP candidates(Black) and freeze-in  
 relic density(Blue) of DM candidate $\chi$ as a function of mass 
 difference  between NLOP and DM candidates and scale of new physics $
 \Lambda$ for Type III Seesaw model with sterile neutrino is shown in the 
 left plot. Lower limit of the lifetime ($10^{-7}$ sec) is coming from the 
  fact that the charged particle has to decay outside the detector whereas 
  the upper limit of 100 sec is coming from BBN  constraints. The current data 
  of CDM relic density put constraints on the  parameter space. The benchmark points 
  we have used for the collider analysis are represented as the black and red points. 
  Right panel shows the yield of the DM candidate (Blue) and NLOP (Red) as a function 
  of  ( $x = \frac{m}{T}$), where $m$ is the mass of the NLOP. We have assumed 
  $M_{\Sigma}$ = 1 TeV and $M_{\nu_{s}}$= 500 GeV. The brown curve shows 
  the equilibrium distribution of the NLOP. The effects after freeze-out of
  NLOP is magnified in the inset.}
 \label{fig:lifetime1}
 \end{center}
 \end{figure}

 In the right panel of Figure~\ref{fig:lifetime1} we depict the evolution of the NLOP (red)
  as well as the DM (blue) as a function of temperature of the Universe. The plot has been generated considering $M_\Sigma$ = 1 TeV, $M_{\nu_s}$= 500 GeV and $\Lambda \sim 10^{14.5}$ GeV.
 The equilibrium yield is shown by the brown curve. We can clearly see that the NLOP depart 
 from the equilibrium and then further decays to the DM depending on its lifetime. 
 The larger the lifetime, later the decay of NLOP and then the yield of NLOP vanishes. 
 For the DM the freeze-in yield increases as the temperature decreases and becomes constant 
 after some time ($x \sim 10 $). The freeze-in yield is dominant when $x\sim 1$ as
 the function $K_1(x)$ which plays the main role in freeze-in production of DM is picked 
 around this value of x. The inset shows that the DM yield gets an additional contribution 
 from the NLOP decay after freeze-out. The relative contribution in the total DM relic density 
 coming from the decay of NLOP depends on the masses of the NLOPs and the DM particle.

 The  decay width of $\eta^{\pm}_3$ into $\chi$ is given by
\begin{equation}
 \Gamma_{\eta^{\pm}_3\rightarrow\chi W^{\pm}} =\frac{g^{2}\sin^{2}\beta \sqrt{E_{w}^{2}-M_{w}^{2}}}
 {4 \pi M_{\Sigma}^{2}}\left(M_{\Sigma}(M_{\Sigma}-E_{w})-3 M_{\nu_{s}}M_{\Sigma} + 
              \frac{2 M_{\Sigma}E_{w}}{M_{w}^{2}}(M_{\Sigma}E_{w}-M_{w}^{2})\right),
\end{equation}
 where $E_{w} = \frac{M_{\Sigma}^{2}-M_{\nu_{s}}^{2}+M_{w}^{2}}{2
  M_{\Sigma}}$. For $M_{\Sigma} = 1$ TeV, $M_{\nu_{s}} = 500$ GeV and $\Lambda = 10^{15}$ GeV the
 lifetime of $\eta^{\pm}_3$ is  0.167 $\small s$.
 For a comparable choice
 of parameters $\psi$ has a lifetime of 0.169 $\small s$ for the decay
 $\psi\rightarrow\chi H$. Thus  $\eta^{\pm}_3$ or $\psi$ never decays inside the 
 LHC detector for such masses and, more importantly, scale of new physics which is 
 at the origin of the dimension-5 terms. The allowed parameter space is explored in
 the left plot of Figure~\ref{fig:lifetime1}. One can thus see
 $\eta^{+}_3\eta^{-}_3$  produced in proton-proton collision via the
 Drell-Yan process, showing up as charged tracks
 all the way up to the muon chamber. The existing mass limit on such
 a quasi-stable particle is 730 GeV from the LHC data till now\cite{CMS:2016ybj}. 
 Since $\Sigma_{3R}$ has zero hypercharge, $\psi$ is only produced in association 
 with $\eta^{\pm}_3$  via W-mediation.  Such final states have the characteristic 
 signature of single heavy stable charged track +  missing transverse energy (MET). 
 In the following  Sections we discuss discovery prospects of
 both of these signals at the LHC.

 An important constraint in this study comes from the light nuclei 
 abundances produced during Big-Bang Nucleosynthesis(BBN). 
 Presence of long lived particles (LLP) may lead to non-thermal 
 nuclear processes (non-thermal BBN) due to their decay into 
 energetic SM particles \cite{Poulin:2015opa,Kawasaki:2004qu,Jedamzik:2006xz}. 
 Non-thermal BBN disturbs the observed light elements abundances which is very 
 close to the estimated abundance according to the standard BBN 
 scenario \cite{Iocco:2008va,Cyburt:2015mya}.
 In our study we have considered an upper bound of the lifetime of $\eta^{\pm}_3$ or $\psi$ to 
 be $\simeq 100$~sec in order to respect the constraint imposed by deuterium 
 abundances during BBN\cite{Banerjee:2016uyt}. Left panel of Figure~\ref{fig:lifetime1} 
 shows variation of lifetime of the NLOP as a function of its mass difference with DM 
 candidate and with the scale of new physics $\Lambda$. It is evident from 
 eqn.~\ref{eq:mixing_beta} that if we increase the new physics scale, the coupling 
 $\sin\beta$, which governs  the decay of NLOP into DM decreases resulting a increase 
 of the lifetime. The light colored region is the allowed parameter space for the Type III 
 seesaw model with sterile neutrinos to spot a HSCP at the LHC. 
 The  blue curves in the left panel of Figure~\ref{fig:lifetime1} shows two different 
 contours of relic density coming from the freeze-in contribution only. As $\Lambda$ 
 increases, the decay width decreases, yielding less DM relic coming from freeze-in 
 production. Since the freeze in contribution can not exceed the total CDM relic density 
 the right side of the right blue contour is disallowed.

\subsection{Inert doublet model (IDM) with right-handed Majorana neutrino}
In this model, the SM particles are postulated to be supplemented with
an additional scalar doublet($\Phi_2$) with hypercharge 1 and three right-handed
SU(2) singlet Majorana neutrinos($N_{iR}$)\cite{Ma:2006km,Chakrabarty:2015yia,Borah:2017dfn}. 
Once more we consider a
$\mathbb{Z}_2$-symmetry to ensure stability of what will emerge as the
DM candidate.  Under the $\mathbb{Z}_2$, two of the Majorana
neutrinos are even and mix with the SM particles to generate neutrino
masses through Type-I seesaw mechanism.  The third Majorana neutrino,
denoted as $N_{3R}$ as well as the additional scalar doublet $\Phi_2$
is  $\mathbb{Z}_2$-odd.
\footnote{Some variants of such a model, postulating all right-handed
Majorana neutrinos to be $\mathbb{Z}_2$-odd, have been studied in 
\cite{Molinaro:2014lfa,Hessler:2016kwm}.} As a result, $\Phi_2$ never 
acquires a vacuum expectation value (vev) and is called the {\em inert 
doublet}.

The scalar potential in this case is
\begin{eqnarray}
V(\Phi_1,\Phi_2) &=& \lambda_1 (\Phi_1^{\dagger}\Phi_1)^2 + \lambda_2
  (\Phi_2^{\dagger}\Phi_2)^2 + \lambda_3
  (\Phi_1^{\dagger}\Phi_1)(\Phi_2^{\dagger}\Phi_2) + \lambda_4
  (\Phi_2^{\dagger}\Phi_1)(\Phi_1^{\dagger}\Phi_2) \nonumber\\
  && + \left[\frac{\lambda_5}{2}(\Phi_1^{\dagger}\Phi_2)^2 + h.c\right]
  +\mu_1\Phi_1^{\dagger}\Phi_1+ \mu_2\Phi_2^{\dagger}\Phi_2,
\end{eqnarray}    
where all parameters are real and $\Phi_1$ is the SM scalar
doublet. The two doublets can be expressed in terms of their components as
\begin{center}
$\Phi_1 = \begin{bmatrix}
G^+ \\
\frac{1}{\sqrt{2}}(v+h+iG^0)
\end{bmatrix}$,$\;$
$\Phi_2 = \begin{bmatrix}
H^+ \\
\frac{1}{\sqrt{2}}(H^0+iA^0)
\end{bmatrix}$,
\end{center}
where, v = 246 GeV, is the electroweak vev. After spontaneous symmetry
breaking, one obtains five physical states $(h,H^0,A^0,H^{\pm})$ and
three Goldstone bosons $(G^0,G^{\pm})$, where $h$ corresponds to the
physical SM-like Higgs field, with mass around 125 GeV.  The $CP$-even 
$(H^0)$, $CP$-odd $(A^0)$ and charged $(H^{\pm})$ scalars 
arise from the inert doublet, since the discrete symmetry
prevents mixing between $\Phi_1$ and $\Phi_2$. The physical scalar masses are
given by,
\begin{subequations}\label{eq:LR-Masses}
\begin{eqnarray}
M_{H^{\pm}}^2 &=& \mu_2 + \frac{1}{2}\lambda_3v^2,\\
M_{H^{0}}^2 &=& \mu_2 + \frac{1}{2}\lambda_Lv^2,\\
M_{A^{0}}^2 &=& \mu_2 + \frac{1}{2}\lambda_Av^2,
\end{eqnarray}
\end{subequations}
where $\lambda_{L/A} = (\lambda_3 + \lambda_4 \pm \lambda_5)$ and
$\lambda_1$ is determined using $M_h$=125 GeV. Note that it is possible
to have substantial mass splittings among $H^0$,$A^0$ and $H^{\pm}$, since 
$\lambda_{3}$, $\lambda_{L}$ and $\lambda_{A}$ are {\it a priori} unrelated.
The scalar potential is bounded from below if it does not turn negative for 
large field values along any possible field direction. In this case, stability 
of the electroweak vacuum is ensured at the electroweak scale and just
above, if the following vacuum stability conditions are satisfied
\cite{Sher:1988mj,Nie:1998yn,Ferreira:2004yd,Branco:2011iw}:
\begin{equation}
\label{vac_sta}
\lambda_1 > 0,\;\;\lambda_2 >
  0,\;\;\lambda_3+\sqrt{\lambda_1\lambda_2} >
  0,\;\;\lambda_3+\lambda_4-|\lambda_5|+\sqrt{\lambda_1\lambda_2} >
  0.
\end{equation}
In addition, we have also ensured that values of the quartic 
interactions used in the phenomenological analyses below are consistent with
the perturbativity bounds, namely, 
\begin{equation}
\label{pert_cond}
\lambda_i < 4\pi , i = 1,\cdots,5. 
\end{equation}

The relevant Yukawa interactions and Majorana mass terms are
\begin{equation}
  \mathcal{L}_Y = y_{\nu j}{\bar {N}}_{3R}{\tilde{\Phi}}_2^{\dagger}L_{Lj} + y_{\alpha
    j}\bar{N}_{\alpha R}{\tilde{\Phi}}_1^{\dagger}L_{Lj} +
  \frac{M_j}{2}\bar{N}_{jR}^c N_{jR} + h.c,
\end{equation}
where $\alpha$ = 1,2 and $j$ = 1,2,3 whereas $L_{L} =
(\nu_L,l_L)^T$. The Yukawa couplings between the $\mathbb{Z}_2$-even
Majorana neutrinos and the SM scalar $\Phi_1$ are responsible for
generation of Dirac type neutrino masses, while the Yukawa coupling
which combines $N_{3R}$ and $\Phi_2$ can only
generate mass for the third neutrino at one-loop level.

If $N_{3R}$ becomes the lightest in the
$\mathbb{Z}_2$-odd sector then the Majorana fermion $\chi = N_{3R} +
N_{3R}^c$ can serve as a viable dark matter candidate. In addition, if
the parameters $\mu_{2}, \lambda_3, \lambda_4, \textrm{ and }
\lambda_5$ are such that $M_{A^{0}} \simeq M_{H^{0}} > M_{H^{\pm}}$ then
the next-to lightest odd particle (NLOP) will be the charged scalar
$H^{\pm}$.

\begin{figure}[t]
    \begin{center}
      \includegraphics[width=7.25cm]{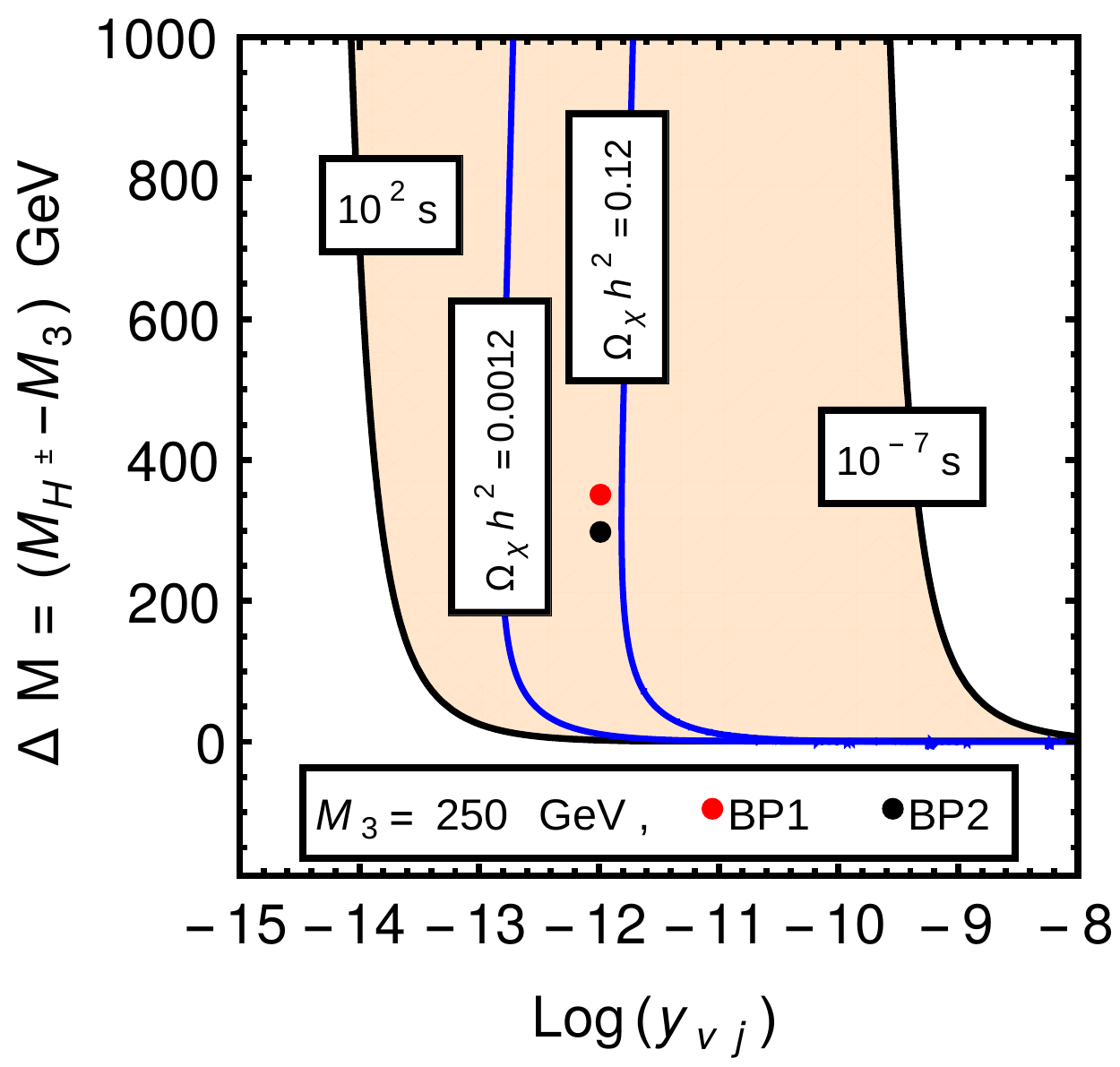}
  \includegraphics[width=7.5cm]{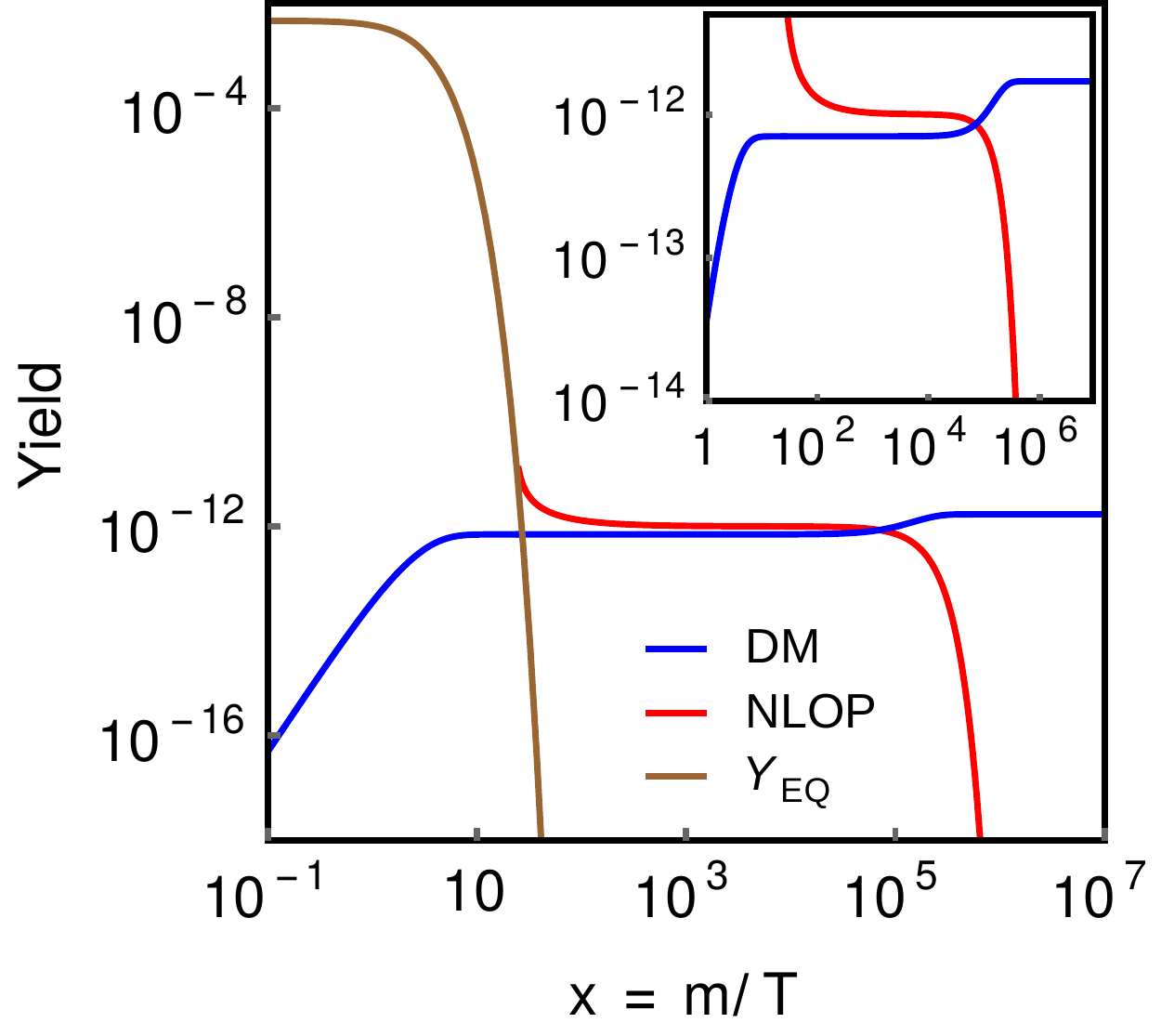}
  \caption{Contours of lifetime of HSCP candidates(Black) and freeze-in 
  relic density of DM candidate $\chi$ (Blue) as a function of mass 
  difference between NLOP and DM candidate and minuscule coupling 
  $y_{\nu j}$ for inert doublet model with  right handed Majorana neutrino
  in the left hand side plot. Detector length restricts the lower limit of lifetime 
  to be $10^{-7}$ sec and the BBN constrains the lifetime to be less than 100~sec.
  Current data of CDM relic density further constrains the parameter space. The 
  benchmark points we have used for the collider analysis are represented as the 
  black and red points. The plot in the right depicts the yield of DM candidate 
  $\chi$ in Blue and that of NLOP in red as a function of x = $\frac{m}{T}$, 
  m being the mass of the NLOP. For this plot
  we have assumed $M_{H^{\pm}}$ = 500 GeV and $M_3$ = 250 GeV. The brown curve shows 
  the equilibrium distribution of the NLOP. The effects after freeze-out of
  NLOP is magnified in the inset.}
  \label{fig:lifetime2}
    \end{center}
\end{figure}

The current neutrino data in principle allow one of the three
light neutrinos to be arbitrarily light. Hence  $y_{\nu j}$ can be 
very tiny (for example, on the order of  $10^{-12}$ while still
being {\em technically natural}). Consequently the DM
candidate $\chi$ will never equilibrate with the thermal
 soup and hence should be treated as a non-thermal DM. 
As we have discussed earlier here also we have computed the DM relic density coming 
from the freeze-in production as well as the later decay of NLOP ($H^\pm$) using
eqns.~\ref{eq:freeze-in-yield} and ~\ref{eq:nlop-decay-yield} where the parameter 
$m_\Sigma$ should be replaced by $m_{H^\pm}$ .

 In the right panel of Figure~\ref{fig:lifetime2} we depict the evolution of the NLOP (red) 
 as well as the DM (blue) as a function of temperature of the Universe. The plot has been generated  considering $M_{H^{\pm}}$ = 500 GeV, $M_{3}$= 250 GeV and $y_{\nu j} \sim 10^{-12}$.
 The qualitative features of the yield of NLOP and DM are same as in the earlier model.
 The plot in the left panel of Figure~\ref{fig:lifetime2} shows the allowed parameter region 
 as a function of $\Delta M$ and $y_{\nu j}$ which is consistent with correct CDM relic density. 
 Also in the left panel of Figure~\ref{fig:lifetime2} we have pointed the benchmark  points 
 for the collider analysis.
 
 The decay width of NLOP into the DM is given by,
 \begin{equation}
 \Gamma_{H^{\pm} \rightarrow \chi l^{\pm}} =\frac{y_{\nu j}^2
   M_{H^{\pm}}}{4
   \pi}\left(1-\frac{M_{3}^{2}}{M_{H^{\pm}}^{2}}\right)^2.
 \end{equation}
 With $y_{\nu j}$ as small as $\simeq 10^{-12}$,
 $M_{H^{\pm}}$ = 500 GeV and $M_3$ = 250 GeV, the lifetime of $H^{\pm}$
 is 0.0297 $s$.  Therefore, for suitable values of parameters as explored in
 Figure~\ref{fig:lifetime2}, $H^\pm$, once
 produced at the LHC  decays outside the detector, leaving its signature
 in the form of a stable charged track.  Since
 $H^{0}$ and $A^{0}$ are heavier than $H^{\pm}$, they may decay into
 $H^{\pm}$ inside the detector depending on the mass splitting. 
 Consequently this scenario can be looked for both 
 opposite-and same-sign heavy stable charged tracks
 ($H^{\pm}H^{\mp},~H^{\pm}H^{\pm}$).
 
In order to be consistent with the recent LHC bounds on long-lived charged 
particles obtained from Drell-Yan production\cite{CMS:2016ybj}, we have always used 
$M_{H^{\pm}} > 360$ GeV. 
%We also find it safe to assume  $M_{H^{0}}\simeq M_{A^{0}} \simeq 
%M_{H^{\pm}}+5$ GeV, so that co-annihilation prevents an inordinately 
%large contribution to the relic density.
As is already mentioned, following the constraint imposed by the light 
element abundances during BBN we have to ensure that the lifetime of our 
proposed LLP candidate  be $\lesssim 100$~sec. The available parameter
space is shown in Figure \ref{fig:lifetime2}, as a function of NLOP mass 
difference with DM candidate and the Yukawa coupling.

%**********************************************************************************************
%----------------------------------------------------------------------------------------------
%**********************************************************************************************

\section{Strategy for analysis}\label{sec:analysis}
% In this  Section we  discuss our strategy for analyzing long lived charged 
% tracks during the future runs of LHC. 
In the collider analysis we have used $\texttt{FeynRules\,2.0}$\cite{Alloul:2013bka} 
and the resulting UFO files are fed into $\texttt{MadGraph5\_aMC@NLO}$\cite{Alwall:2014hca} to 
generate our Signal events. Parton showering as well as hadronization is done 
using  $\texttt{Pythia\,6}$\cite{Sjostrand:2006za}. Finally the detector simulation is done using 
$\texttt{Delphes\,3}$\cite{deFavereau:2013fsa} framework. For the signal generation we have used 
$\texttt{CTEQ6L1}$\cite{Pumplin:2002vw} as our Parton Distribution Function. 
%The relic density is calculated using $\texttt{MicrOmegas\,4.2.5}$\cite{Belanger:2014vza}.

The characteristic features of these heavy stable charged tracks is that the tracks have large
transverse momenta which distinguish them from muons which also are charged and stable in the
collider scale. Moreover, since these particles are substantially heavy (500 GeV or more) these
charged tracks behave just like a {\em slow} muon, i.e. their velocity $\beta = p/E$ is considerably
lower than unity, which is not the case for muons. 
Such slow charged tracks of massive NLOP will have
high specific ionization rate $\frac{dE}{dx}$, and 
are delayed in their flight between the tracker and the muon chamber.
 The CMS and ATLAS collaborations\cite{Chatrchyan:2013oca,ATLAS:2014fka}
have already used these high ionization properties and time of flight measurements
 to distinguish them from the muons. In this work we put cuts on transverse momentum 
and $\beta$ of the heavy charged tracks to distinguish them 
from background muons. The exact values of the cuts that are used in our analysis
are given in Table \ref{cut1}.

\begin{table}[h]
\begin{center}
\begin{tabular}{|c|c|c|c|c|}
\hline
Parameter &$\beta$& $p_T$ &$|y(\mu_{1,2})|$ &$\Delta R(\mu_1 ,\mu_2)$ \\\hline\hline
Cut values&(A)[0.2, 0.95] &$> 70$ GeV&$< 2.5$ & $> 0.4$ \\
          &(B)[0.2, 0.80]&$> 70$ GeV&$< 2.5$ & $> 0.4$ \\ \hline
\end{tabular}
\caption{Basic selection cuts applied to analyze signals of heavy stable 
charged track.}
\label{cut1}
\end{center}
\end{table}

Cut set (A) above corresponds exactly to the ATLAS specification
\cite{ATLAS:2014fka}. In cut set (B), we have experimented a bit by 
inserting a stronger $\beta$-cut following \cite{Chatrchyan:2013oca}, while 
keeping everything else unchanged. This stronger $\beta$-cut is somewhat
more effective, since it removes all backgrounds which retaining enough
signal events even at low luminosity. This is, especially true for `single 
charged track events' studied later.

$\beta$ being an important observable in our analysis, one requires 
realistic velocity distribution for both our charged tracks and muons, 
the most important candidate for our background. 
The velocity distribution of the muons from a combined measurement of the 
calorimeter and muon chamber has a small spread with mean 
$\bar{\beta}=0.999c$ and a standard deviation $\sigma_{\beta}=0.024c$
\cite{ATLAS:2014fka}. Hence in our analysis we generate a Gaussian random 
number with these parameters and then impose the cuts on the smeared $\beta
$ accordingly. For the heavy charged particles we have used $\bar{\beta}=p/
E$ and the same value for $\sigma_{\beta}$ as above. 

Next, we discuss the proposed benchmark points and backgrounds considered 
for each of the respective channels.

\subsection{Type III seesaw with sterile neutrino}\label{BP-typeIII}

In this model we focus on the following channels, viz.
\begin{itemize}
\item Opposite-sign charge tracks: $$p\;p \to Z^* \to \eta^{\pm}_3\;\eta^{\mp}_3.$$ 
\item Single charge track + $\met$: $$p\;p\to W^{\pm *}\to\eta^{\pm}_3\;\psi.$$
\end{itemize} 
We have chosen the benchmark points given in Table \ref{benchmark1}, which is well inside the
available parameter space as explored in Figure \ref{fig:lifetime1}.
We have also set  the Wilson coefficients 
$\alpha_{\Sigma \nu_s}$, $\alpha_{\Sigma}$ and $\alpha_{\nu_s}$ to unity,
as already stated.

\begin{table}[h]
\begin{center}
\begin{tabular}{|c|c|c|c|}\hline
  Parameters &  $M_{\Sigma}$ (GeV) & $M_{\nu_{s}}$ (GeV) & $\Lambda$ (GeV)  \\\hline\hline
  BP1        & 850 & 500 &$10^{15}$ \\\hline
  BP2        & 950 & 500 & $10^{15}$\\\hline
\end{tabular}
\end{center}
\caption{Benchmark points for studying the discovery prospects of stable 
charged tracks of $\eta^{\pm}_{3}$ and $\psi$ for Type III seesaw model at 14 TeV run of LHC.}
\label{benchmark1}
\end{table}

%need to rephrase 
In Figure~\ref{fig:cross_section} we have plotted production cross-section of 
$\eta^\pm_3\,\eta^\mp_3$ and $\eta^\pm_3\,\psi$ at 14 TeV LHC. The cross-section of 
$\eta^\pm_3\,\psi$ is larger due to the coupling with the $W$-boson.  
  
\subsubsection{Background Estimation}\label{sec:typeIII_BG}
The SM background for opposite-sign charged tracks 
is muons coming mainly from Drell-Yan production of $\mu^{\pm}$, $\tau^{\pm}$
(computed at NNLO)\cite{Catani:2009sm} and $t\,\bar t$ (N$^3$LO)\cite{Muselli:2015kba}. 
We have also  considered the sub-dominant backgrounds coming from $W^+ W^-$, $WZ$ and $ZZ$ 
final states (NLO)\cite{Campbell:2011bn}.

For analyzing the signal of single heavy stable charged track + $\met$,  
$W^\pm$ (NNLO)\cite{Catani:2009sm} and $t\bar t$ (N$^3$LO)\cite{Muselli:2015kba}
final states are the dominant backgrounds. Diboson ($W^+ W^-$, $WZ$ and $ZZ
$)\cite{Campbell:2011bn} final states are sub-dominant 
backgrounds which, too, have taken into account.
% To perform our analysis, we generated a statistically significant number 
%of  events such that we are sure of the number of events after putting the 
%cut  given in Table \ref{cut1}. 

In order to be as realistic as possible, we have also 
considered the background coming from cosmic ray muons.
Following the analysis of CMS, cosmic ray muon constitutes nearly 60$\%$ of
total background in case of opposite-sign heavy charged tracks
\cite{Chatrchyan:2013oca}. 
For single charged track+$\met$ analysis, due to lack of available
information in the literature, we have assumed the cosmic ray muon 
backgrounds is the same as in the case of opposite-sign charge tracks.
However, even if we assume such background to be one order of magnitude 
larger than that in case of opposite-sign charged track pairs the net background
cross-section changes by about 0.5$\%$ only. This is because the SM 
backgrounds arising from LHC processes dominates over the cosmic ray muon
backgrounds, when it comes to a single observed tracks. Based on this observation,
we believe that our background estimate is convergent and realistic.

The opposite sign dimuon background after the selection cuts (A) of 
Table~\ref{cut1} is 2.667 $fb$ while the single muon track + $\met$ has a 
background cross-section 3368.6 $fb$. The single muon track + $\met$ 
background can be  further reduced by applying 
a suitable $\met$ cut as discussed later in Section~\ref{single_track_met}.

\subsection{Inert doublet model with right-handed Majorana neutrino}
\begin{figure}[t]
    \begin{center}
  \includegraphics[width=10cm]{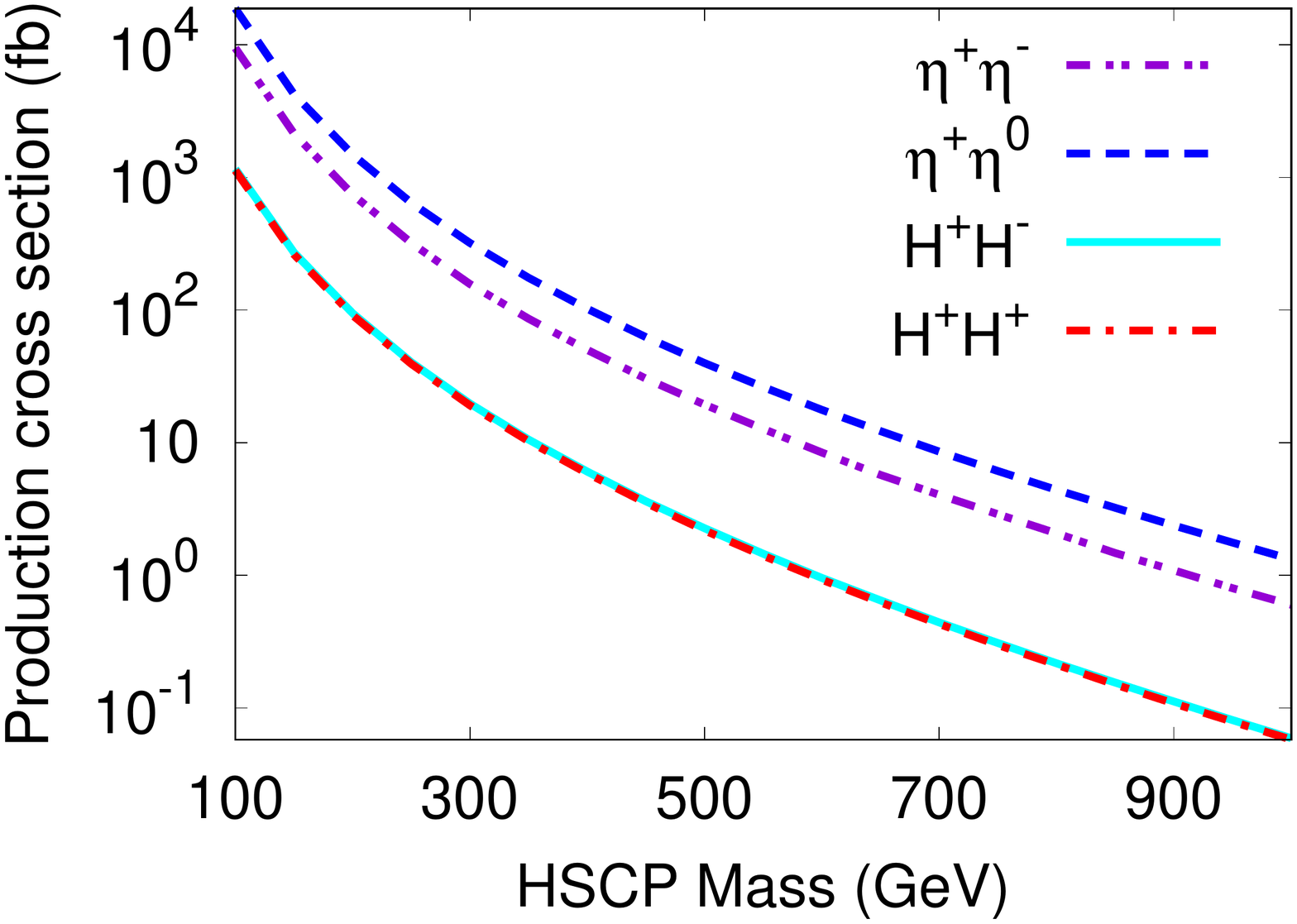}
  \caption{Production cross-section of heavy stable charged particles 
  (HSCP) of Type III Seesaw model and IDM  at the 14TeV LHC. Here dashed blue line 
  also includes the production cross-section of $\eta^{-}\eta^{0}$ and dot-dashed
  red line shows the cross-section for $H^{\pm}~H^{\pm}$. }
  \label{fig:cross_section}
    \end{center}
\end{figure}

We have explored the following signals in this scenario : 
\begin{itemize}
\item Opposite-sign charge tracks ($ H^\pm H^\mp$)
\item Same-sign charge tracks ($ H^\pm H^\pm$)
\end{itemize}
The second channel is possible here because $H^0(A^0)$, being a self-conjugate
particle, can decay into $H^{+}$ and $H^{-}$ with equal probabilities. 
This is not expected in the previously considered scenario with
$\eta^{\pm}_3$ and $\psi$ being nearly-degenerate. The dominant production channel for these signals is the following 
\be
p  p \to W^{\pm *} \to H^\pm\,H^0 \to H^\pm (H^\pm W^{*\mp} / H^\mp W^{*\pm}).
\ee

\begin{table}[ht]
\begin{center}
\begin{tabular}{|c|c|c|c|c|c|c|c|}\hline
  Parameters &  $M_{H^{\pm}}$(GeV) & $M_{H^{0}}$(GeV) & $M_{A^{0}}$(GeV)&$M_{3}$ (GeV) & $\lambda_{2}$  &$\lambda_{L}$ &$y_{\nu j}$  \\\hline\hline
  BP1        & 550 & 555 & 555 & 250 & 0.5 & 0.04 & $10^{-12}$ \\\hline
  BP2        & 600 & 605 & 605 & 250 & 0.5 & 0.04 &  $10^{-12}$  \\\hline
\end{tabular}
\end{center}
\caption{Benchmark points for studying the discovery prospects of stable 
charged tracks of $H^{\pm}$ for IDM  at 14 TeV run of LHC.}
\label{benchmark2}
\end{table}

The production cross-sections of signal processes in the 14 TeV run of the LHC are
shown in Figure~\ref{fig:cross_section}.
The opposite-sign charge track production also gets additional contribution from 
$Z$-mediation which is order of magnitude smaller than the dominant channel.

The benchmark  points that are used in our analysis are tabulated 
in Table~\ref{benchmark2} are all allowed according to Figure~\ref{fig:lifetime2},
and also satisfy vacuum stability and perturbativity criteria 
given in eqns.~\ref{vac_sta} and eqn.~\ref{pert_cond}.

\subsubsection{Background Estimation}
For the signal corresponding to two opposite-sign heavy charged tracks 
we have considered the same backgrounds as is already discussed in 
Section \ref{sec:typeIII_BG}.

In case of the same-sign heavy stable charged tracks the dominant 
backgrounds (same sign dimuons) are coming from $t\bar t$ (N$^3$LO)
\cite{Muselli:2015kba}, $t\bar t W$ (NLO)\cite{Campbell:2012dh} and diboson 
final states (NLO)\cite{Campbell:2011bn}. We have also considered the sub-dominant 
backgrounds like $W^{\pm}\,\gamma$ and $WW +2jets$. 
Cosmic ray muon background is considered to be the same as in 
the case of opposite-sign charged tracks in order to be conservative enough
regarding background estimation.
%%%%%%%%%%%%%%%%%%%%%%%%%%%%%%%%%%%%%%%%%%%%%%%%%%%%%%%%%%%%%%%%%%%%%%%%%%%%%%%%%%%%%%%%%%%%%%%

\section{Results and Discussions}\label{sec:result}

In this  Section we have discussed the discovery prospects of heavy stable charged tracks
in the considered benchmark points during 14 TeV runs of LHC. We compute the 
statistical significance of the proposed final states using the standard formula
\begin{equation}\label{eq:significance}
\mathcal{S} = \frac{N_S}{\sqrt{N_S + N_B}},
\end{equation}
where $N_S$ and $N_B$ are respectively number of signal and background 
events passing the cuts.
\subsection{Type III seesaw with sterile neutrino}

\subsubsection{Opposite-sign stable charged tracks of $\eta^{\pm}_3\,\eta^{\mp}_3$} 

\begin{figure}[t]
    \begin{center}
  \includegraphics[width=5cm,angle=270]{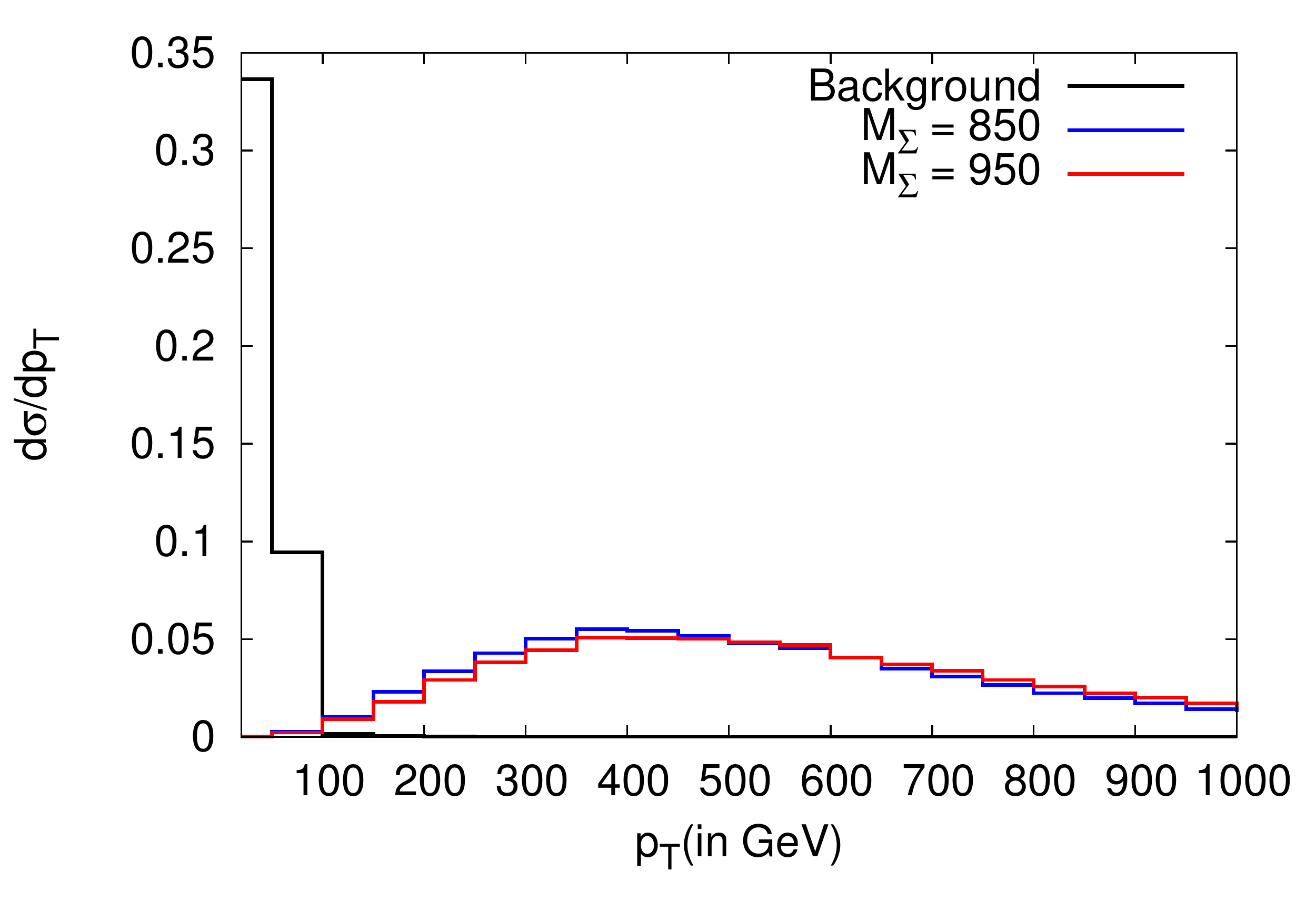}
  \includegraphics[width=5cm,angle=270]{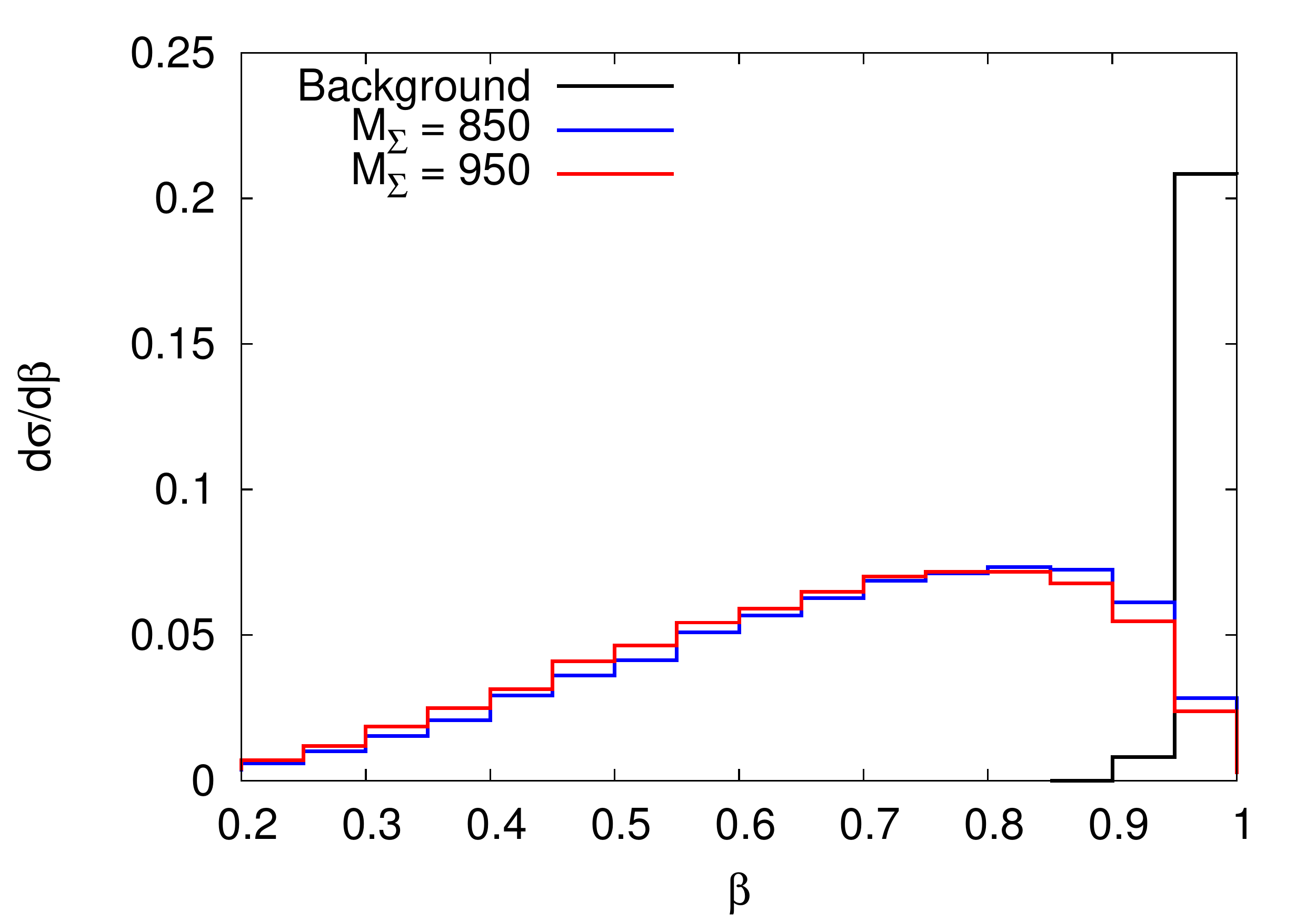}
  \caption{$p_T$-and $\beta$-distribution of the opposite sign stable charged tracks of
    $\eta^{\pm}_{3}\,\eta^{\mp}_{3}$ for Type III seesaw with sterile neutrino for the 
    benchmark points BP1(blue) and BP2(red) 
    as in Table \ref{benchmark1}. Background  muon distribution is shown in black histogram.}
  \label{fig:seesaw_spectrum}
    \end{center}
\end{figure}

\begin{table}[t]
\begin{center}
\begin{tabular}{|c|c|c|c|c|c|}
\hline
Signal & Benchmark point & $\int\!\mathcal{L}dt$ for 5$\sigma$ & $N_S$ & $N_B$ & $N_S/N_B$ \\ [0.5ex]
\hline \hline
\multirow{2}{*}{\parbox[t]{3cm}{Opposite Sign \\ Charged Track }} 
           & BP1 & 92.95  & 92  & 248  &  0.37\\\cline{2-6}
           & BP2 & 263.23 & 146 & 702  &  0.21\\
\hline\hline
\multirow{2}{*}{\parbox[t]{3cm}{Single Charged \\ Track + $\met$ }} 
           & BP1 &  (A)340.40  &  841  & 27436 & 0.030  \\
           &     &  (B) 24.81  &   46  &    40 & 1.150  \\\cline{2-6}
           & BP2 &  (A)1076.19 &  1485 & 86741 & 0.017  \\ 
           &     &  (B)  56.60 &   62  &    91 & 0.681  \\
\hline
\end{tabular}
\caption{Integrated luminosity ($fb^{-1}$)required to attain 5$\sigma$ statistical significance for 
  opposite sign charged tracks and single charged track  +$ \met$ signals for the considered 
  Benchmark points of Table~\ref{benchmark1} in the Type III seesaw with a sterile neutrino model during 14 TeV run of LHC.}
\label{tab:reqd-lum-typeIII}
\end{center}
\end{table}

We have presented in Figure~\ref{fig:seesaw_spectrum} the $p_T$-and $\beta$-distribution 
of the harder charged track for the two benchmark points BP1(blue) and BP2(red) 
as in Table~\ref{benchmark1}. We have 
also shown the background dimuon distribution in solid black histogram.
The signal tracks tend to have higher $p_{T}$ owing to the NLOP mass.
At the same time, the fact that they are produced by Drell-Yan process
close to kinematic threshold in the parton center-of-mass frame endows
them with $\beta$ well below unity. Thus one is able to distinguish NLOP
tracks from muons using the $p_{T}$-and $\beta$-cuts listed in Table~\ref{cut1}.
The imposition of such cuts allows one to predict a statistical significance of
5$\sigma$ for various integrated luminosities, as listed in Table~\ref{tab:reqd-lum-typeIII}.

\begin{figure}[t]
    \begin{center}
  \includegraphics[width=7cm,angle=270]{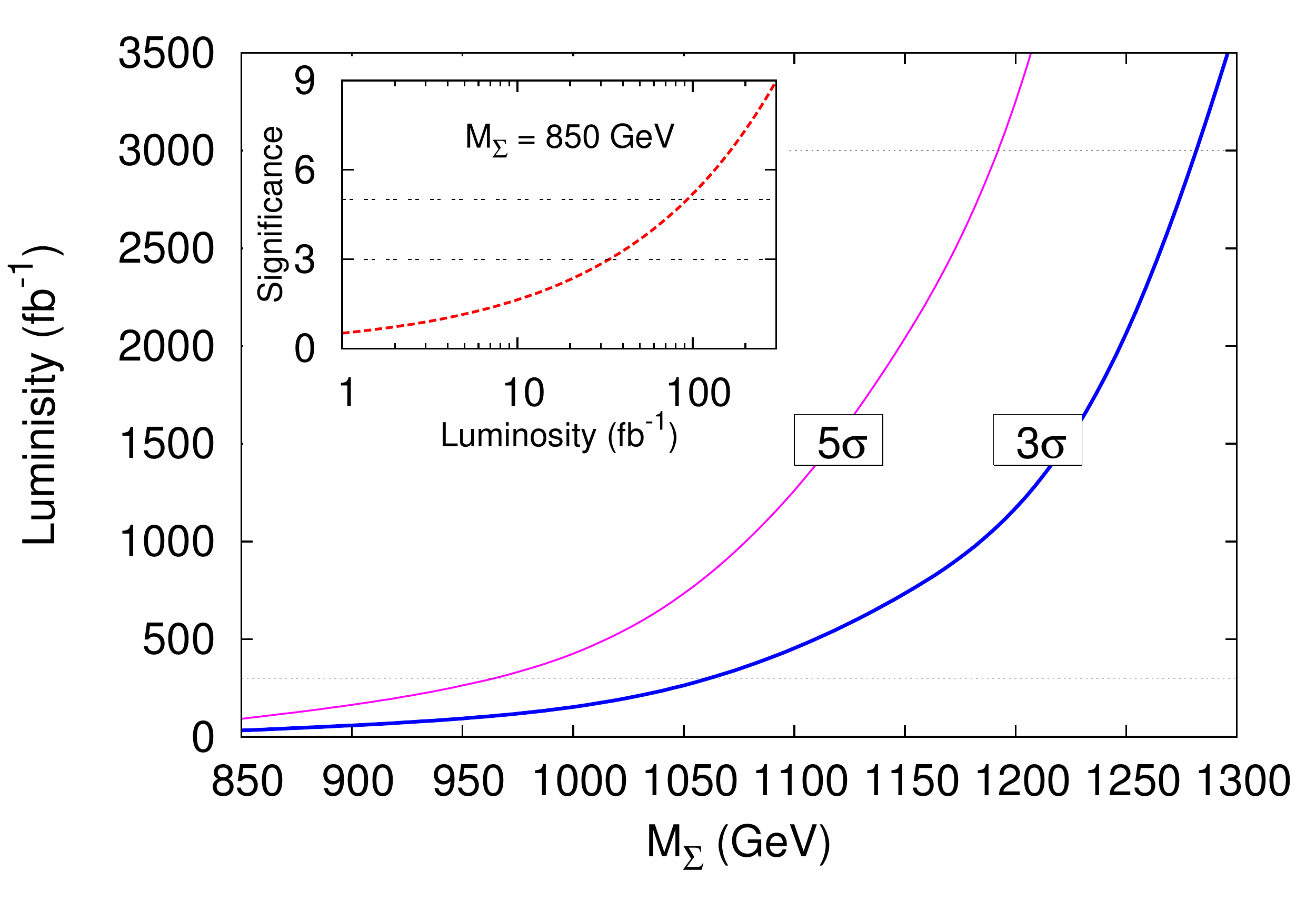}
  \caption{Projection of $3\sigma$(blue) and $5\sigma$(magenta) significance contours 
  as a function of HSCP mass and integrated luminosity for the Type III seesaw with 
  a sterile neutrino. Inset depicts the significance of BP1 with integrated luminosity 
  during 14 TeV runs of LHC.}  
  \label{fig:seesaw_lum_mass_reach}
    \end{center}
\end{figure}

In Figure \ref{fig:seesaw_lum_mass_reach} we have shown $3\sigma$(blue) 
and $5\sigma$(magenta) significance contours at the 14 TeV LHC, in terms of heavy 
charged particle and integrated luminosity. The horizontal lines represent 
integrated luminosities of 300~$fb^{-1}$ and 3000~$fb^{-1}$.  As we can see with 
the 14 TeV run of LHC this model can be probed up to $ M_\Sigma =  1060(960)$ GeV
with $3\sigma(5\sigma)$ significance with integrated luminosity of $300 fb^{-1}$.
Whereas at HL-LHC with $3000 fb^{-1}$ data the model can be explored up to 
$\sim 1280(1190)$ GeV. The inset in Figure~\ref{fig:seesaw_lum_mass_reach} depicts the projected
significance for the BP1  as a function of integrated luminosity. For BP1  
$3 \,\sigma(5 \,\sigma)$ significance can be achieved 
with integrated luminosity $\sim 35(100) fb^{-1}$.

%*********************************************************************************************
\subsubsection{Single stable charged tracks of $\eta^{\pm}_3$ + $\met$}
\label{single_track_met}
\begin{figure}[t]
    \begin{center}
  \includegraphics[width=7cm,angle=270]{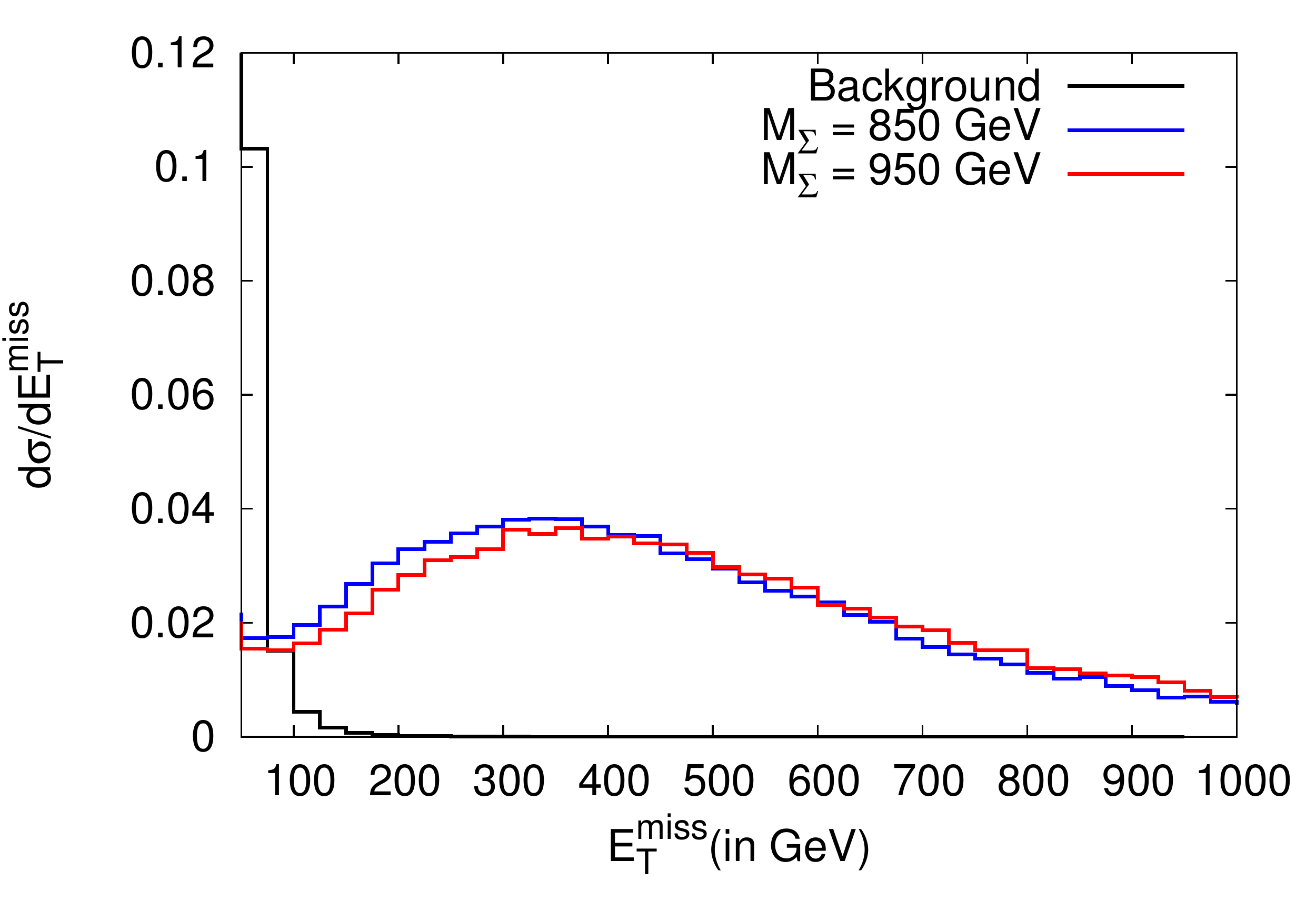}
  \caption{$\met$ distribution of the single heavy stable charged tracks of
    $\eta^{\pm}_{3} + \met$ for the benchmark points BP1(blue) and BP2(red) 
    as in Table \ref{benchmark1}. Background  $\met$ distribution is shown 
    in black.}
  \label{fig:seesaw_MET_spectrum}
    \end{center}
\end{figure}

The $\eta^{\pm}_3$ charged track + $\met$ also has appreciable
production cross-section and at the same time suffers from a large background from $W^{\pm}$ 
production at the LHC\cite{TheATLAScollaboration:2015xzd}.
One obviously has to go beyond the basic cuts listed in Table \ref{cut1} in 
order to size down the background efficiently. However, one has an 
additional handle in the form of large $\met$, since the production process is 
$p p \rightarrow \eta^{\pm}_{3} \psi$, and $\psi$ is a massive neutral fermion whose
decay rate is again suppressed by $\frac{1}{\Lambda^{2}}$. Thus we have
put an additional cut $\met > $  150 GeV to reduce the background substantially. 
The $\met$ distribution for background and signal 
events are shown in Figure~\ref{fig:seesaw_MET_spectrum}. After putting the 
$\met$ cut along with cut set A of Table \ref{cut1} the background cross-section reduces to 80.6 $fb$.

The required integrated luminosities to reach 5$\sigma$ statistical significance 
for this signal(using cut set (A)) during 14 TeV run of LHC for each of the benchmark points is tabulated
in Table \ref{tab:reqd-lum-typeIII}. If in addition we consider the cosmic ray muon
background to be one order of magnitude larger than that with dimuons, even then one obtain 
5$\sigma$ statistical significance for BP1(BP2) with an integrated luminosity 343(1080) 
$fb^{-1}$. The difference is small because the SM background at LHC is dominant. 

The production cross section for single charged track +$\met$ is large compared to the 
opposite-sign charged track in Type III seesaw with sterile neutrino model. However, the huge 
single muon SM background pushes towards higher luminosities to reach 5$\sigma$ statistical 
significance compared to the opposite sign stable charge track signal.
If we impose the cut-set (B), which applies a more stringent cut
on $\beta$, together with a $\met$-cut of 150 GeV,
the SM background can be reduced enormously. Thus one can probe this signal
at a much lower value of integrated luminosity as shown in Table~
\ref{tab:reqd-lum-typeIII}.

However, although cut-set (B) eliminates the SM background completely in all 
other cases,
the results do not improve much, as the SM background in those cases is 
already small enough 
after the imposition of cut (A).

%*********************************************************************************************

\subsection{Inert doublet model with right-handed Majorana neutrino}

\subsubsection{ Opposite-sign stable charged tracks of $H^{\pm}H^{\mp}$}
 The $H^{\pm}$ particles are massive and  the strong $p_T$-and $\beta$-cuts as listed in 
Table \ref{cut1} are quite effective in reducing the SM backgrounds drastically. Hence the lion's
share of the background contribution comes from the  cosmic ray muons. The required integrated
luminosities for 5$\sigma$ statistical significance for the considered benchmark points is shown
in Table \ref{tab:reqd-lum-IDM}.

\begin{table}[h]
\begin{center}
\begin{tabular}{|c|c|c|c|c|c|}
  \hline
  Signal & Benchmark point & $\int\!\mathcal{L}dt$ for 5$\sigma$ & $N_S$ & $N_B$ & $N_S/N_B$ \\ [0.5ex]
\hline \hline
\multirow{2}{*}{\parbox[t]{3cm}{Opposite Sign \\ Charged Track }} 
           & BP1 &  97.81 &  94 & 261 & 0.36 \\\cline{2-6}
           & BP2 & 195.16 & 127 & 520 & 0.24 \\
\hline\hline
\multirow{2}{*}{\parbox[t]{3cm}{Same  Sign \\ Charged Track}} 
           & BP1 &  71.62  & 67 & 115 & 0.58  \\\cline{2-6}
           & BP2 & 137.45  & 88 & 220 & 0.40  \\  
\hline
\end{tabular}
\caption{Integrated luminosity($fb^{-1}$) required to attain 5$\sigma$ statistical significance for $H^{\pm}\,H^{\mp}$ 
signal for the considered benchmark points during 14 TeV run of LHC.}
\label{tab:reqd-lum-IDM}
\end{center}
\end{table}
 
\begin{figure}[t]
    \begin{center}
  \includegraphics[width=7cm,angle=270]{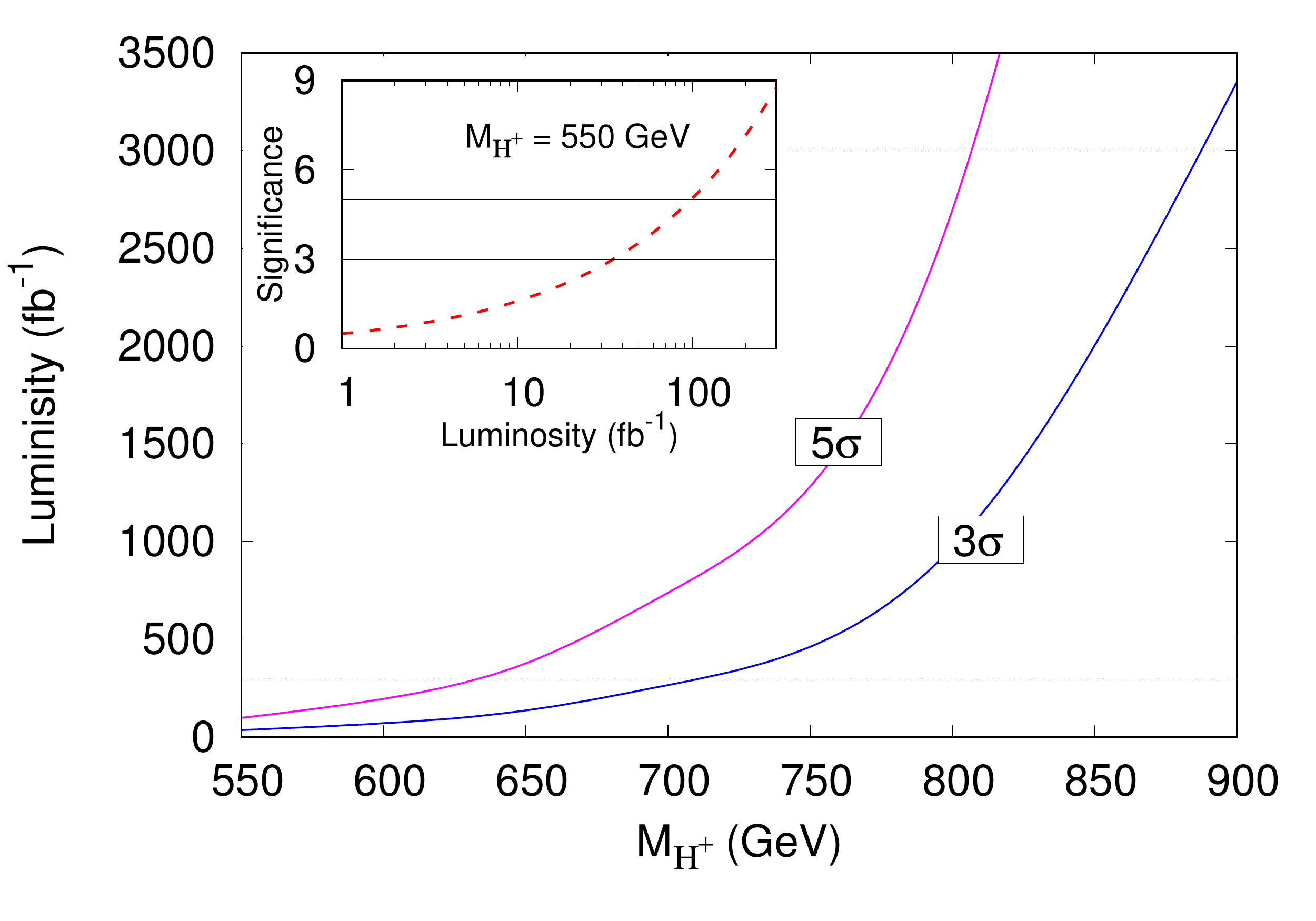}
  \caption{Projection of $3\sigma$(blue) and $5\sigma$(magenta) significance contours 
    for the opposite-sign charged tracks as a function of HSCP mass and integrated 
    luminosity for the IDM with a RH 
  Majorana neutrino. Inset depicts the significance of BP1 with integrated 
  luminosity during 14 TeV runs of LHC.}
  \label{fig:IDM_lum_mass_reach}
    \end{center}
\end{figure}

The required integrated luminosities to obtain 5$\sigma$(magenta) and 3$\sigma$(blue) 
statistical significance for different values of $M_{H^{\pm}}$ is shown in 
figure \ref{fig:IDM_lum_mass_reach}. Clearly one can see that with 3000(300)$fb^{-1}$ of
integrated luminosity this model can be tested up to $M_{H^{\pm}}$ = 880(720) GeV with
$3\,\sigma$ significance. The $5\,\sigma$ discovery limit for this model is 
$M_{H^{\pm}}$ = 800(630) GeV with integrated luminosity of 3000(300) $fb^{-1}$. 
The slightly lower reach compared to the previous case can be attributed to the
lack of enhancement via polarisation sum, when it comes to the production of the 
quasi-stable charged scalar. In spite of this small degree of suppression, it is 
clear that here, too, the high energy run of the LHC should reveal signals of 
such a scenario (as well as the previous one discussed here) even before the 
high luminosity run begins. 

\subsubsection{ Same-sign stable charged tracks of $H^{\pm}\,H^{\pm}$}
\begin{figure}[t]
    \begin{center}
  \includegraphics[width=7cm,angle=270]{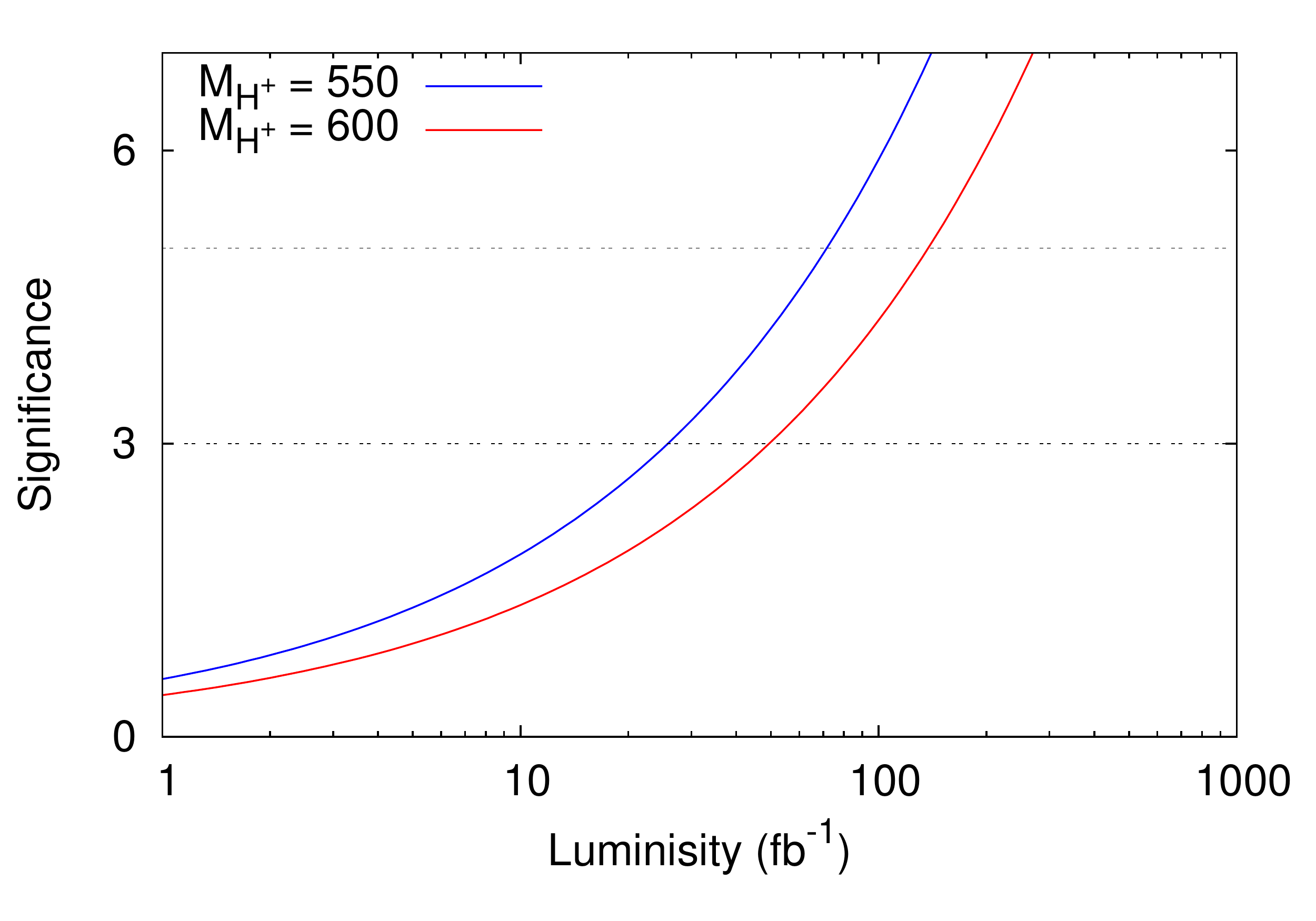}
  \caption{Variation of statistical significance of the considered 
  benchmark points with integrated luminosity($fb^{-1}$) for same-sign double 
  charged tracks.}
  \label{fig:IDM_BP_lum}
    \end{center}
\end{figure}

While single track events are not expected here because of the short-lived
nature of $H^{0}(A^{0})$, one anticipates same-sign charged track pairs in this
case, as has been mentioned already. Here too, the $p_T$-and $\beta$-cuts (of the
values (A)) suffice to suppress all SM background. However, we have also
introduced a third-muon veto to further reduce such background. Since
there is no Drell-Yan production of same-sign dimuons, these criteria 
eliminates the SM background almost completely, leaving 
only cosmic ray muons. Once more, we have assumed the same rate as in the 
case of opposite sign charged track pairs.
The required integrated luminosity for 5$\sigma$ statistical significance 
for the considered benchmark points is shown in Table \ref{tab:reqd-lum-IDM}.
The figures in that table indicate that the luminosity requirements for 
same-and opposite-sign tracks are comparable. This is because $p p
\rightarrow H^{0}(A^{0})H^{\pm}$
yields both type of track pairs. While the latter signal has additional 
contributions
from s-channel $H^{+}H^{-}$ production, the SM backgrounds for opposite 
track
pairs are also more copious. This is mainly because large mass 
splittings between charged and neutral higgses is prohibited by the 
requirement of perturbative unitarity of scalar quartic couplings and also 
the fact that $\mathbb{Z}_2$ symmetry prevents mixing between two higgs 
doublets. Thus we have only a minor excess of opposite-sign 
track pair events. The variation of statistical significance of this signal 
with integrated luminosity is shown in Figure \ref{fig:IDM_BP_lum}.

The same-sign charged track pairs is a unique feature of this model 
and largely depends on the mass of heavy scalar and pseudoscalar. We found 
that for $M_{H^{\pm}} = 550$ GeV perturbative unitarity dictates 
$M_{H^{0}/A^{0}} \simeq 800$ GeV  and the production cross-section for 
same sign track is $0.2\;fb$ and can be discovered at 5$\sigma$ with 
an integrated luminosity of 2500 $\rm{fb}^{-1}$. Beyond this mass one 
has to rely on opposite-sign charged tracks to search for the IDM model
with  a RH Majorana neutrino with superweak interaction.

\section{Summary and conclusions}\label{sec:Summary and conclusions}
 There are situations when the DM is feebly interacting 
 and is consequently non-thermal . In this scenario the usual MET signal is 
 not present and if the next heavier particle which decays to the DM is 
 electrically charged then we can see long lived charged tracks at the 
 detector due to its long lifetime compared to the detector scale. 

 Here we have presented two non-supersymmetric models of feebly interacting 
 DM. These are a variant of Type III seesaw with a sterile neutrino and 
 the inert doublet model (IDM) with at least one right-handed Majorana 
 neutrino. We have analyzed and constrained the parameter space where the 
 DM candidate is a  {\it SuperWIMP} dark matter. In both the models the 
 next-to-lightest-particle is electrically  charged and long lived which 
 decays to DM to yield correct 
 CDM relic  density. The lifetime of these NLOPs are large enough to pass 
 through the  detector but decays within $100$~sec and do not disturb the 
 BBN. 
 
 We have identified a few benchmark points within the constrained model 
 parameter space and studied the collider prospect of the stable charged particles 
 at the 14 TeV run of the LHC. For the Type III  seesaw with sterile neutrino model
 we have considered the opposite sign  charged track and also single charge track with
 large $\met$. The IDM with right-handed Majorana neutrino model
 gives rise to both same-sign and opposite-sign charge tracks.
 These `alternative' subsidiary signals may be helpful
 in differentiating between two theoretical frameworks, both of which admit
 a stable charged NLOP.
 
  We have presented the discovery potential for two benchmark points 
  for each scenario: the Type-III case allows to probe higher mass ranges
  because of the enhancement of production cross-section via fermion polarization
  summation. We find that the  extension of Type III  Seesaw including a sterile neutrino 
  can be probed about 960(1190)GeV with 300(3000) $fb^{-1}$, whereas the IDM with 
  right-handed Majorana neutrino model can be explored up to 630(800)GeV. 
  It should also be noted that the above results use leading order
  production rates for the NLOP, while the background rates used have
  take into account higher order enhancements (NLO/NNLO/N$^3$LO). Thus the
  search limits predicted are on the conservative side, and one may expect
  even more  optimistic results, once  higher-order contributions to the signal(s)
  are included \cite{Ruiz:2015zca,Fuks:2017vtl,Muhlleitner:2003me}. 
 
%****************************************************************
\section{Acknowledgements}\label{sec:Acknowledgements}
We thank Shankha Banerjee and Anirban Biswas for helpful discussions.
This work was partially supported by funding available from
the Department of Atomic Energy, Government of India, for the Regional Centre for
Accelerator-based Particle Physics (RECAPP), Harish-Chandra Research Institute. 
%****************************************************************
%++++++++++++++++++++++++++++++++++++++++++++++++++++++++++++++++++++++++++++++++++++++++++++++++++++++++++++++
\appendix
\section{Mixing between charged fermions} \label{app:charged-mixing}
In  this appendix we illustrate how the charged component of the triplet fermion 
in our first model (with $\mathbb{Z}_2$ odd fermion triplet and RH neutrino) can be 
made lighter than the neutral component. Let us introduce vector-like SU(2) singlet singly 
charged weyl fermions $\lambda_{L,R}$ and a triplet scalar $\Delta$ with $Y=2$. 
If $\lambda_{L,R}$ are odd and $\Delta$ is even under the imposed $\mathbb{Z}_2$ 
symmetry then the relevant part of the Lagrangian is 
\be\label{eq:app-mass-matrix}
\mathcal{L} = -M_\lambda \bar{\lambda} \lambda 
	      - (Y_\lambda\; \textrm{Tr}\left[\bar{\Sigma}_{3R}^c \;\Delta\; \lambda_R + 
	      \bar{\Sigma}_{3R} \;\Delta\; \lambda_{R}^{c}\right] + h.c), 
\ee
where ${\Sigma}_{3R}$ is defined in eqn.~\ref{eq:sigma-bi-doublet},
 $\lambda = \lambda_{L} + \lambda_{R}$ and $\Delta$ is defined as 
\be\label{eq:delta-bi-doublet}
\Delta=\begin{bmatrix}
  \delta^+/\sqrt2 & \delta^{++}\\
  \delta^0 & -\delta^+/\sqrt2.
       \end{bmatrix}
\ee
Once the neutral component of the triplet scalar acquires a vev $v_\Delta$ the Yukawa term of the above Lagrangian will generate 
a mixing between $\lambda_{L,R}$ and $\eta_3^\pm$ and the charged fermion mass matrix will become 
\be
\mathcal{M}^\pm = \begin{bmatrix}
                   M_\Sigma-\frac{\alpha_{\Sigma}v^2}{2\Lambda}&v_\Delta Y_\lambda\\
		   v_\Delta Y_\lambda^\dagger & M_\lambda
                  \end{bmatrix},
\ee
where $M_\Sigma$ is the mass of the fermion triplet as given in eqn.~\ref{eq:lag_ren} and 
$Y_\lambda$ is taken to be real for simplicity.  
The vev of the triplet scalar is restricted by the experimental observation of the $\rho$ parameter and 
we assume $v_\Delta = 4$ GeV which is well within the current limit~\cite{Arhrib:2011uy}. The eigenvalues of the mass matrix (eqn.~\ref{eq:app-mass-matrix}) will be(neglecting the tiny contribution of higher-dimensional term) 
\be
\dfrac12 \left[M_\Sigma+M_\lambda\pm\sqrt{\left(M_\Sigma-M_\lambda\right)^2+4 v_\Delta^2\;Y_\lambda^2}  \right]. 
\ee
From the above equation it is evident that if $M_\lambda > M_\Sigma$ then the lightest state  will be 
triplet dominant with a mass slightly smaller than $M_\Sigma$ which for all practical purposes, can be identified 
as  $\eta_3^\pm$. In Table.~\ref{tab:app-table} we have tabulated the exact eigenvalues for a few benchmark points. 
The masses of the triplet fermion $M_\Sigma$ are kept at the same values as those used in our phenomenological 
analysis. We fix the illustrate Yukawa coupling to a value consistent with perturbativity. 
Evidently the mixing between charged  fermions pulls down the mass of 
$\eta_3^\pm$ by about 250 MeV or more from $M_\Sigma$ depending on the 
mass of the heavy vector-like fermion. This offsets the upward revision by appropiately 166 MeV via electromagnetic corrections, as given, 
for example in~\cite{Cirelli:2005uq}. On the other hand, the neutral component of $\Sigma_3$ mixes only 
with the $\nu_s$ via tiny dimension-five operators and its mass will remain at $M_\Sigma$. Hence, the mass of $\eta_3^\pm$ remains below the 
neutral component mass for all the benchmark points shown in Table~\ref{tab:app-table}
and hence explains the long-lived nature of $\eta_3^\pm$.

The extent to which $\eta_3^\pm$ is lighter than $\psi$ depends on the Yukawa coupling, $Y_\lambda$ and the triplet vev.
However this formulation demonstrates that having the neutral fermion above, or degenerate with, its charged 
SU(2) partner is not inconceivable. Moreover, depending on the splitting  between $M_\psi$ and $M_{\eta_3^\pm}$
 one can have an additional signal, either single-track events or two-track ones with a displaced vertex for 
one of the tracks. Results for the first case alone have been presented in the text.  
% Moreover, one should note that we can play around with $Y_\lambda$ and bring down $\eta_3^\pm$ mass exactly 
% by 166 MeV so that the charged and neutral component of fermion triplet are degenerate.

% \usepackage{multirow}
% \usepackage{multirow}
\begin{table}[]
\centering

\renewcommand{\arraystretch}{1.35}
\begin{tabular}{|c|c|c|c|c|}
\hline
\multirow{2}{*}{\begin{tabular}[c]{@{}c@{}}$M_\Sigma$\\ (GeV)\\$\approx M_\psi$\end{tabular}} & \multirow{2}{*}{\begin{tabular}[c]{@{}c@{}}$M_\lambda$\\ (GeV)\end{tabular}} & \multirow{2}{*}{$Y_\lambda$} & \multicolumn{2}{c|}{Eigenvalues}                                                                            \\ \cline{4-5} 
&  &   & \begin{tabular}[c]{@{}c@{}}Light(GeV)\\$\approx M_{\eta_3^\pm}$\end{tabular} & \begin{tabular}[c]{@{}c@{}}Heavy (GeV)\end{tabular} \\ \hline\hline
\multirow{2}{*}{850}          & 2000      & 5       & 849.65         & 2000.35      \\ \cline{2-5} 
                              & 2500      & 5       & 849.76         & 2500.24      \\ \hline
\multirow{2}{*}{950}          & 2000      & 5       & 949.62         & 2000.38     \\ \cline{2-5} 
                              & 2500      & 5       & 949.74        & 2500.26      \\ \hline
\end{tabular}
\caption{Eigenvalues of the nearly degenerate charged and neutral fermions for few benchmark points after mixing between the triplet fermion and 
vector-like heavy charged fermion. }
\label{tab:app-table}
\end{table}

\providecommand{\href}[2]{#2}
\addcontentsline{toc}{section}{References}
\bibliographystyle{JHEP}
\bibliography{nonsusy}

\end{document}